\documentclass[12pt,a4paper]{article} 
\usepackage{graphicx}
\usepackage{amsmath}
\usepackage{subfig}
\usepackage{makecell}
\usepackage{ulem,slashbox}
\usepackage{cite}

\begin{document}

\title{What do the cosmological supernova data really tell us?}
\renewcommand{\thefootnote}{\fnsymbol{footnote}}
\author{\. Ibrahim Semiz\footnote{e-mail: ibrahim.semiz@boun.edu.tr} ~\& A. Kaz\i m \c Caml\i bel\footnote{e-mail: kazim.camlibel@boun.edu.tr}
\\ \small Physics Department, Bo\u gazi\c ci University, \\ \small Bebek, \. Istanbul, Turkey}
\date{ }

\maketitle
\renewcommand{\thefootnote}{\arabic{footnote}}\setcounter{footnote}{0}
\begin{abstract}

Not much by themselves, aparently.

We try to reconstruct the scale factor $a(t)$ of the universe from the SNe Ia data, i.e. the luminosity distance $d_{L}(z)$, using only the cosmological principle and the assumption that gravitation is governed by a metric theory. In our hence {\it model-independent}, or {\it cosmographic} study, we fit functions to $d_{L}(z)$ rather than $a(t)$, since $d_{L}(z)$ is what is measured. We find that the acceleration history of the universe cannot be reliably determined in this approach due to the irregularity and parametrization-dependence of the results.

However, adding the GRB data to the dataset cures most of the irregularities, at the cost of compromising the model-independent nature of the study slightly. Then we can determine the redshift of transition to cosmic acceleration as $z_{\rm t} \sim 0.50 \pm 0.09$ for a flat universe (larger for positive spatial curvature).

If Einstein gravity (GR) is assumed, we find a redshift at which the density of the universe predicted from the $d_{L}(z)$ data is independent of curvature. We use this point to derive an upper limit on matter density, hence a lower limit on the density of dark energy. While these limits do not improve the generally accepted ones, they are derived {\it only using the $d_{L}(z)$ data.}

\end{abstract}


\section{Introduction}

Acceleration in the expansion rate of the universe was first realised at the very end of the $20^{\rm{th}}$ century with the SNe Ia observations~\cite{supernovateam, supernovaproject}. Since then, the cause of this phenomenon has become one of the most prominent questions in cosmology. Among the mainstream explanations are the Cosmological Constant $\Lambda$, scalar field Dark Energy (quintessence~\cite{quintessence}, phantom~\cite{phantom1}, etc.), modified gravity~\cite{modified} and braneworld scenarios~\cite{dgp}. 

How {\it do} these observations tell us about the acceleration of the universe? They consist of measurements of redshift $z$ and distance modulus $\mu$ for each supernova. Since the type Ia supernovae are taken as standard candles~\cite{SN1a_std_cdl}, the distance moduli can be converted to luminosity distances $d_L$. One interprets them in the framework of the FRW cosmological models, which follow from the assumptions of homogeneity and isotropy --the celebrated cosmological principle introduced by Einstein and amply verified by the high degree of isotropy of the cosmic background radiation-- and leads to a description of the spacetime metric of the universe in terms of a single function $a(t)$, the {\it scale factor}. In these models, the relation between the redshift and scale factor  is given by
\begin{equation}
a(t)=\frac{a_0}{1+z(t)}. \label{scalefactorredshift}
\end{equation}
where $a_{0}\equiv a(t_{0})$, $t_{0}$ denoting the age of the universe . Then the luminosity distance of a source at redshift $z$ is determined by the expansion history of the universe as~\cite{SN1a_std_cdl}: 
\begin{equation}
d_L(z)=(1+z)a_{0} f^{-1}\left( \frac{c\int_{0}^{z}\frac{dz}{H(z')}}{a_{0}}\right)
\label{eq:lumdisfroma}
\end{equation} 
where $f(r)$ is given\footnote{In eq.(\ref{eq:lumdisfroma}), the large round parenthesis contains the argument of the function $f^{-1}$; it is not a factor multiplying the expression $f^{-1}$.} by 
\begin{equation}
f(r)=c\int_{t}^{t_0}\frac{dt'}{a(t')}=-\int_{r}^{0}\frac{dr'}{\sqrt{1-kr'^2} }
\end{equation}
with $k$ being the curvature parameter which takes values ($-1, 0, 1$) for spatially open, flat or closed universes respectively, and $H(t)=\dot{a}(t)/a(t)$ the Hubble parameter. Note that these results  follow from the assumption of the FRW metric, without using Einstein or alternative theory of gravity.

Simply put, the observations found supernovae at a given redshift $z$ to be {\it fainter} than what would be expected even for an empty flat universe, i.e. for one with $a(t) \propto t$, that is, with zero acceleration/deceleration. This faintness puts the sources farther away as compared to the empty universe, hence farther back in time, due to the constant speed of light. On the other hand, by eq.(\ref{scalefactorredshift}) the redshift gives the size of the universe at the emission of the light, hence the faintness pushes the bottom of the $a$~vs.~$t$ curve to the left, hence acceleration. 

The original works~\cite{supernovateam, supernovaproject} used the $\Lambda$CDM model, i.e. assumed that the universe is dominated by matter [corresponding to a perfect fluid with equation of state (EoS) $p=0$; and some of it possibly dark, although this does not matter for the Einstein equations], since we know that matter exists in the universe, and a cosmological constant (corresponding to a perfect fluid with EoS $p=-\rho$) as the agent causing the acceleration. They parametrized the contributions of matter and cosmological constant as fractions of the so-called {\it critical density},  called $\Omega_{m}$ and $\Omega_{\Lambda}$, and performed fits of the data to the $d_{L}(z)$ functions calculated for assumed values of these parameters. On the $\Omega_{m}$-$\Omega_{\Lambda}$ plane, the best fits they found suggested that $\Omega_{\Lambda}-\Omega_{m} \simeq 0.4$, consistent with an accelerating universe. The best fits give very weak constraints on the sum $\Omega_{\Lambda}+\Omega_{m}$ determining the curvature of the universe, however; that quantity was first meaningfully constrained by the measurement of the location of the first peak in the power spectrum of the fluctuations in the cosmic background radiation~\cite{cmb_first_peak}. 

Of course, other models have been put forward since, e.g. featuring fluids with different EoS's. To evaluate any such theory, or determine the best values of the theory's parameters, usually the same procedure is followed; i.e. $a(t)$ is determined from the assumptions of the theory, then $d_{L}(z)$ is evaluated using eq.(\ref{eq:lumdisfroma}), and checked against the data to assess the viability of the model.

However, the procedure of going from a predicted $a(t)$ to a predicted $d_{L}(z)$ can in principle be inverted. We can solve for $t(z)$ from eq.(\ref{eq:lumdisfroma}):
\begin{equation}
t_{0}-t= \int_{0}^{z} \frac{dz}{c(1+z)}\frac{1}{\sqrt{1-\kappa\frac{d_L^2(z)}{(1+z)^2}}}\frac{d}{dz}\left[ \frac{d_L(z)}{1+z}\right],                \label{lookback}
\end{equation}
where the curvature $\kappa$ is defined by
\begin{equation}
\kappa =\frac{k}{a_0^2}.   \label{kappa}
\end{equation}
As mentioned above, we keep the $\kappa$-term, since the SNe Ia data do not tell us that the universe is flat (or otherwise). Once $t(z)$ is determined, this function can in principle be inverted to give $z(t)$.

Hence, if we have good knowledge of the function $d_{L}(z)$ \underline{and} if the integration (\ref{lookback}) and subsequent inversion can be performed analytically, we would have a good candidate for an analytical expression for the expansion history of the universe via eq.(\ref{scalefactorredshift}). Then the derivatives of that function would reveal the history of the Hubble parameter, and periods of acceleration or deceleration.

Since no theory of gravity (Einstein or modified) or assumption about the content of the universe is used, this is called a {\it model-independent approach}, or since the emphasis is on simply determining the unknown metric function of the universe, occasionally {\it cosmography}~\cite{cosmography}.

In this work, we follow this approach. We fit families of functions to the $d_{L}(z)$ data, and determine the best fitting member of each family. The choice of function families is not motivated by a physical model we have in mind, rather by possible simplicity of the analytical operations. Trying to find good fits to $d_{L}(z)$ rather than to $a(t)$, or some combination of its derivatives such as the deceleration parameter $q=a\ddot{a}/\dot{a}^{2}$, etc. is better-motivated, since the observationally measured function is $d_{L}(z)$.  We also use alternative redshift variables in addition to the conventional redshift $z$, introducing one of our own, and denoting them by $y_{i}$. We compare goodness of fit of the results with each other, and in particular, with the standard model of cosmology, the $\Lambda$CDM model. We find that none of the models we investigated fits meaningfully better than $\Lambda$CDM.

Then we attempt to extract all the possible information about the expansion history of the universe, including periods of acceleration and deceleration. It turns out that not much analytical progress can be achieved in terms of the time $t$, but both the first and second time-derivatives of the scale function $a(t)$ can be evaluated {\it in terms of a redshift variable $y_{i}$} analytically, once an analytical result for $d_{L}(y_{i})$ is assumed. We find that while the SNe Ia data \textit{strongly suggest} that the present acceleration of the universe is positive, they \textit{by themselves} cannot tell us that the universe was decelerating in the past. However, we find that including the GRB luminosity-distance data improves the situation, even though they are much fewer in number,  subject to much larger uncertainties, and not \textit{really} model-independent. We also rederive the result that {\it for given $d_{L}(y_{i})$}, the present value of acceleration of the universe is independent of its curvature (of course, for FRW models), contrary to some claims in the literature (e.g. \cite[Figure 1]{mortsellclarkson}).

When we finally \textit{do} use Einstein's Equations to derive conclusions about the content of the universe in the past according to General Relativity, we find that there is a special value of redshift at which the density of the universe is independent of its curvature \textit{for a given $d_{L}(y_{i})$.} We  use this fact to put an upper limit on the matter density of the universe, leading to a lower limit on dark energy. We also investigate the evolution of the total EoS parameter, all analyses performed \textit{using only the SN+GRB data}. 
 

\section{Fitting the luminosity distance function}

\subsection{Alternative variables} \label{sect:redshiftvars}

The conventional redshift $z$ is not the only possible redshift variable, not necessarily even the best variable to work with. Cosmography has usually been the effort to extract as many Taylor expansion coefficients as possible of various cosmological functions as functions of time or redshift; for example, the Hubble and deceleration parameters as first and second terms. But one look at eq.(\ref{scalefactorredshift}) shows that the radius of convergence of the Taylor series for $a(z)$ is 1, hence some such Taylor series will not be reliable beyond $z=1$~\cite{cosmography}; and alternative redshift variables have been used in the literature \cite{cosmography, yvariables}.  We will use these variables too, and introduce one of our own.

We will occasionally denote the conventional redshift $z$ by $y_{0}$. The first alternative variable is~\cite{cosmography}
\begin{equation}
y_{1}=\frac{z}{z+1},
\label{yredshift}
\end{equation}  
other alternatives introduced in \cite{yvariables} are
\begin{align}
& y_{2}=\arctan\frac{z}{z+1}, & y_{3}=\frac{z}{z^2+1}, && y_{4}=\arctan{z}, &
\label{othery}
\end{align}  
we introduce\footnote{During the writing of this manuscript, we became aware of another work~\cite{ln(1+z)} introducing the same variable.}
\begin{equation}
y_{5}= \ln{(z+1)}.
\label{logredshift}
\end{equation}  
and also will use the almost-trivial $y_{6}=u=z+1$.

\begin{figure}[h!]
\caption{\textsf{Relation betweeen the redshift variables $y_{0}$-$y_{5}$ and the scale factor. Note that $y_{3}$ is not monotonic with $a$.}} 
\centering
\includegraphics[width=0.9 \columnwidth]{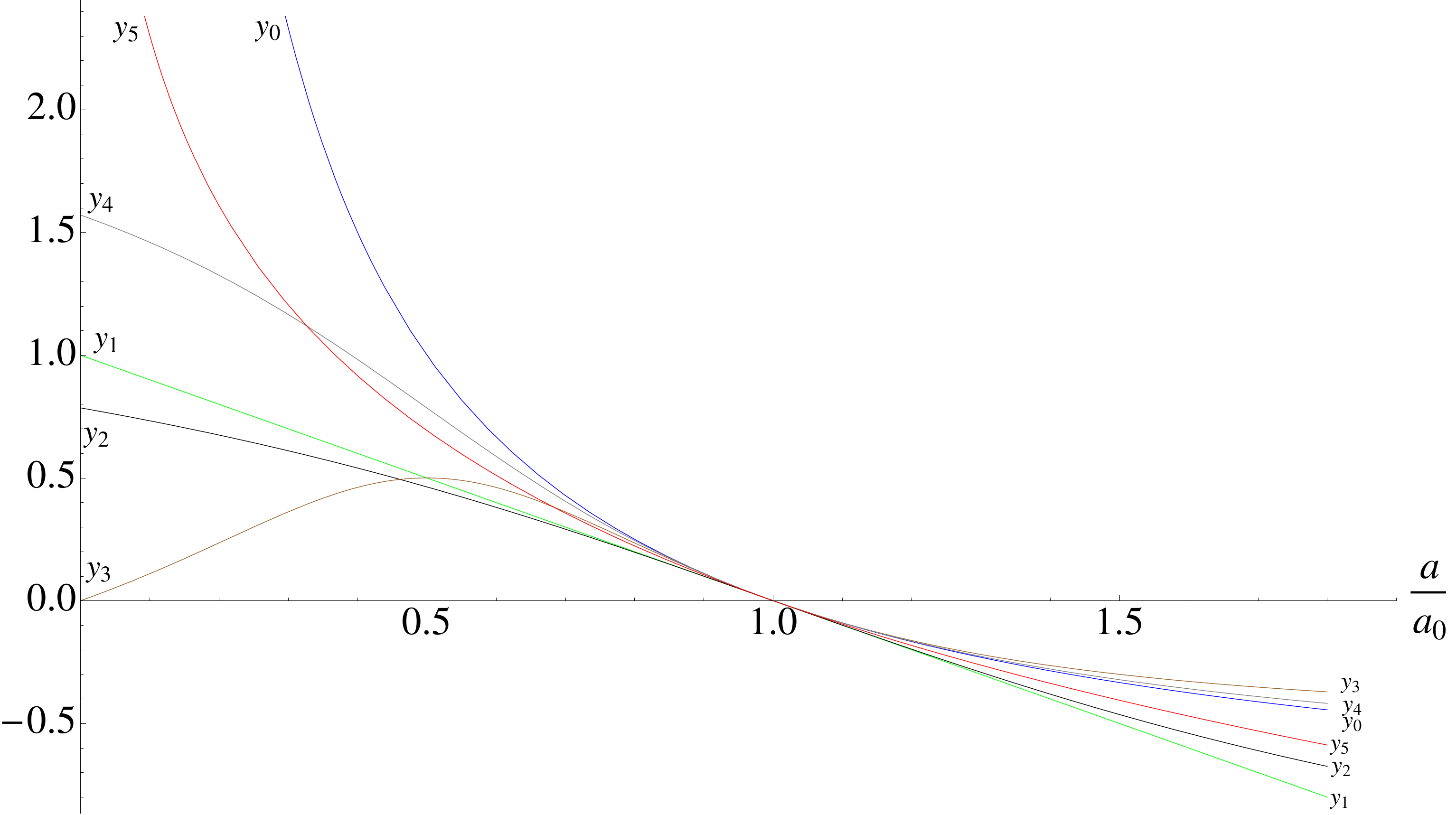}
\label{redshifts} 
\end{figure}

Behaviours of the different redshift variables as function of the scale factor are shown in Fig \ref{redshifts}. It can be seen that $y_{3}$ does not change monotonically with $z$, therefore $a(t)$, hence we will not use that variable.

\subsection{The candidate families and the fits} \label{familiesANDfits}

To determine candidate functions for $d_L(y_{i})$, we choose 7 families of functions (See Table \ref{introducingfamilies}) to fit to the data, that is, the Union 2.1 dataset~\cite{union21}, with $\chi^2$ minimisation. To be as generic as possible we start with polynomials with no constant term [except when using $y_{6}=u=z+1$, when the constant term is determined in terms of the coefficients of the other powers such that $d_{L}(u=1)=0$]. The second and third families are polynomials multiplied by an exponential function where the exponent is linear and quadratic in $y_{i}$, respectively. We extend the spectrum by introducing three more families which are the first three families multiplied by $1+z$; this is done to simplify (\ref{lookback}) where division of $d_L$ by $1+z$ is present. Of course, when an alternative redshift variable is used, this factor is expressed in terms of that variable, e.g. $(1-y_{1})^{-1}$. One further generalisation of polynomials is the Pade approximant \cite{pade1, pade2, pade3} which is given by 
\begin{equation}
{\rm Pade} (y, M, N)=\frac{P_M(y_{i})}{1+P_N(y_{i})}
\end{equation}
where $P_M(y_{i})$ is the $M^{\rm th}$ order polynomial with constant term set to zero; and we have added one family consisting of Pade approximants multiplied by $(1+z)$. 
\begin{table} [h!]
\caption{\textsf{The 8 different families used in fits. $y$ can be any one of the redshift parameters described in Sect. \ref{sect:redshiftvars},  $P_N(y)$ is the $N^{\rm th}$ order polynomial with zero constant term (except when using $y_{6}=u$, see beginning of Sect. \ref{familiesANDfits}), $u(y)$ is $(1+z)$ expressed in terms of $y$, Pade($y$, $M$, $N$) is the Pade approximant in variable $y$ and orders $M$ \& $N$; and $c$ \& $d$ are constants. }}
\centering
\begin{tabular}{| c | c |}
\hline
Designation & Function family  \\
\hline
F1 & $P_N(y)$  \\
\hline
F2 & $P_N(y) u(y)$  \\
\hline
F3 & $P_N(y) \exp(c y)$   \\
\hline
F4 & $P_N(y) u(y) \exp(c y) $   \\
\hline
F5 & $P_N(y) \exp(c y + d y^2)$   \\
\hline
F6 & $P_N(y) u(y) \exp(c y + d y^2) $   \\
\hline
F7 & Pade($y$, $M$, $N$)   \\
\hline
F8 & $u(y)$ Pade($y$, $M$, $N$)   \\
\hline
\end{tabular}
\label{introducingfamilies}
\end{table} 

\begin{figure} [h!]
\caption{\textsf{Union 2.1 data (in terms of luminosity distance and standard redshift $z \equiv y_{0}$); the N=2 to 7 fits for the first family, i.e. simple polynomials; the fits or the MD (black) and $\Lambda$CDM (red) models, and the one-sigma confidence-levels of the best-fitting member (N=4) of the family F1. The inset shows the right end, magnified.}}
\centering
\includegraphics[width=0.99 \columnwidth]{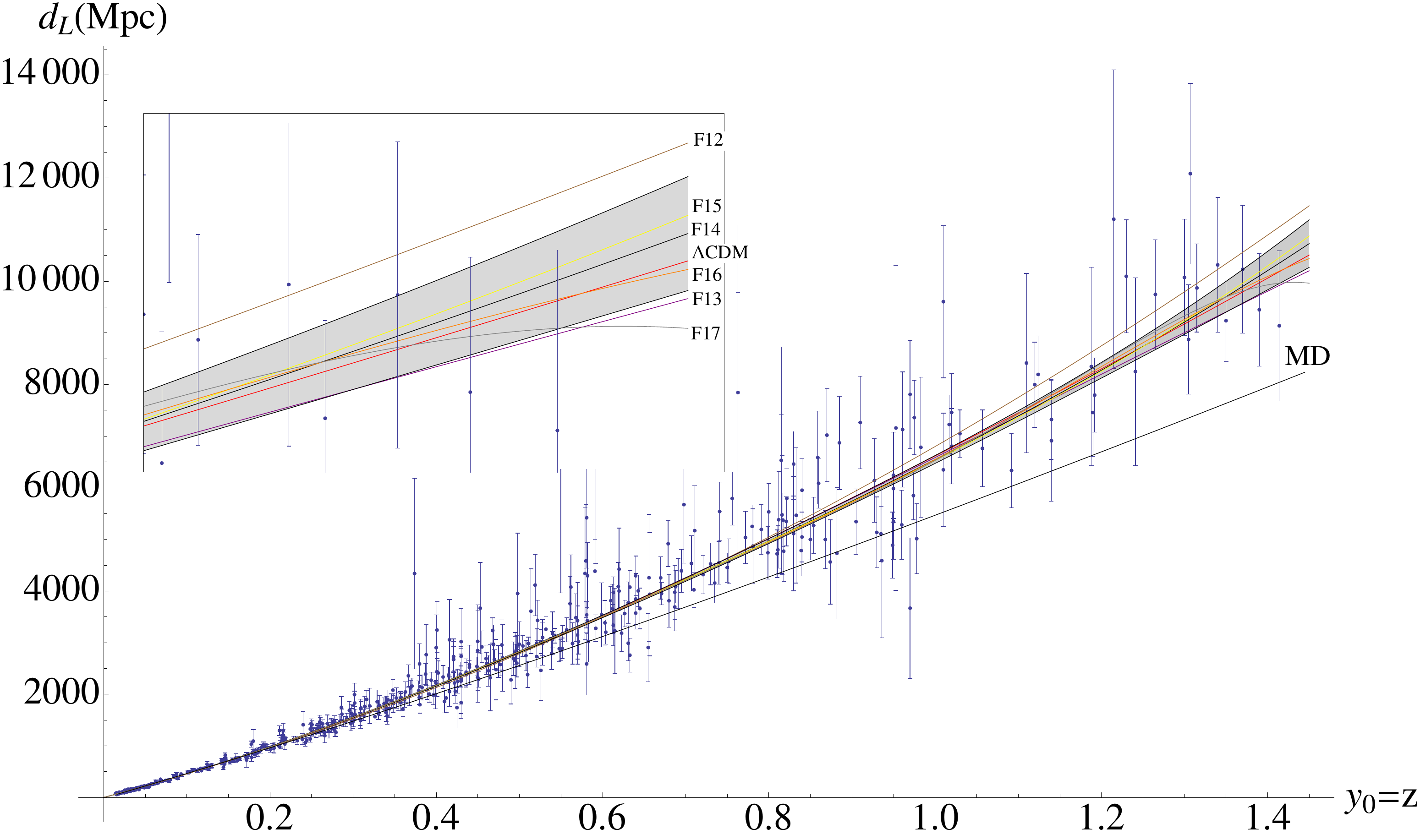}
\label{fig:polymodels-y0} 
\end{figure}

For each family, we determine the best-fitting member, i.e. the one with the lowest $\chi^2_\nu \equiv \chi^2$/d.o.f value. As an illustration, in Fig.\ref{fig:polymodels-y0} we show the Union 2.1 data (in terms of luminosity distance and standard redshift $z \equiv y_{0}$); the N=2 to 7 fits for the first family, i.e. simple polynomials; the curves for the flat matter-dominated model of the 1990's (based on the $H_{0}$ value derived from the data) and the best-fitting $\Lambda$CDM model; and the one-sigma confidence-levels of the best-fitting member of the family F1. We can see that the matter-dominated model is comfortably excluded. The best-fitting polynomial is the $4^{\rm th}$ order one, telling us that no more than four independent cosmological parameters can be meaningfully extracted from the Union 2.1 data, if one expands the function $d_{l}(z)$ into a MacLaurin series. 

\begin{figure}[b!]
\caption{\textsf{Union 2.1 data (in terms of luminosity distance and the redshift $z \equiv y_{0}$; and the best fits for each function family listed in Table \ref{families}. The one-sigma confidence-levels are not shown in order not to clutter up the figure; they are similar to those in Fig.\ref{fig:polymodels-y0}.The inset shows the right end, magnified.}}
\centering
\includegraphics[width=0.99 \columnwidth]{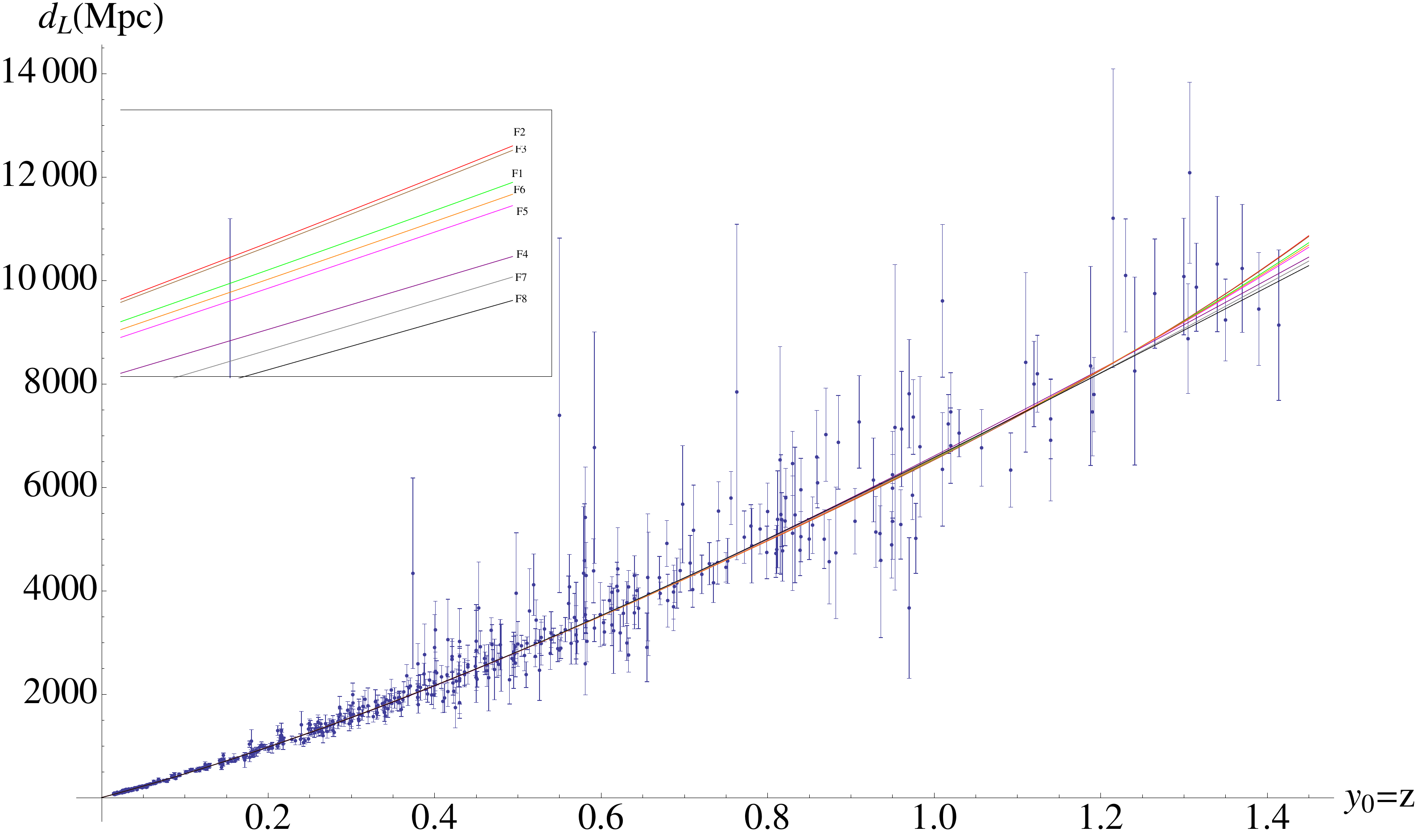}
\label{fig:bestfits-y0} 
\end{figure}

\begin{figure}[h!]
\caption{\textsf{Union 2.1 data (in terms of luminosity distance and the redshift  variable $y_{1}$); and the best fits for each function family listed in Table \ref{families}. The one-sigma confidence-levels are not shown in order not to clutter up the figure; they are similar to those in Fig.\ref{fig:polymodels-y0}. The inset shows the right end, magnified.}}
\centering
\includegraphics[width=0.88 \columnwidth]{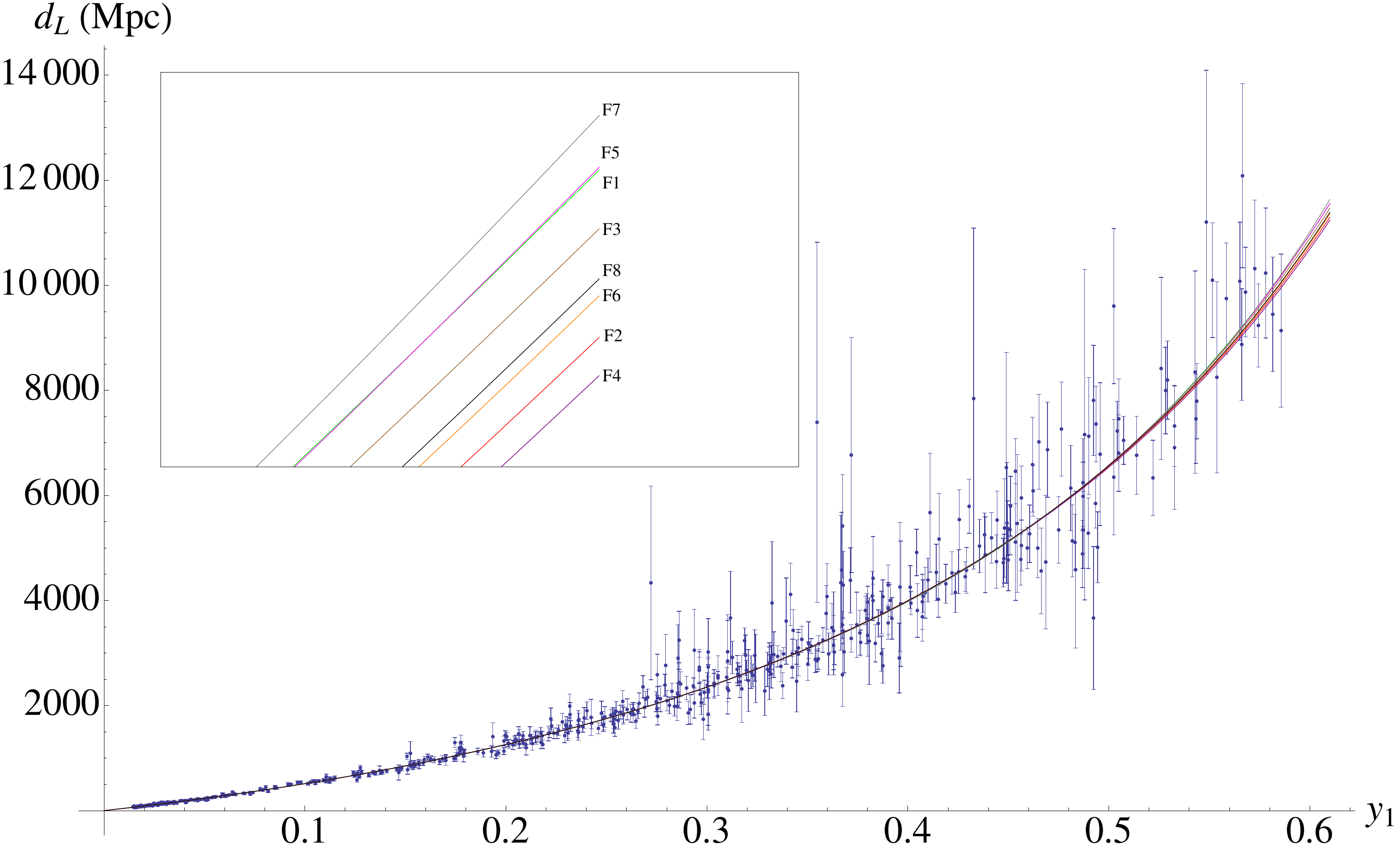}
\label{fig:bestfits-y1} 
\end{figure}

\begin{figure}[h!]
\caption{\textsf{Union 2.1 data (in terms of luminosity distance and the redshift variable $y_{2}$; and the best fits for each function family listed in Table \ref{families}. The one-sigma confidence-levels are not shown in order not to clutter up the figure; they are similar to those in Fig.\ref{fig:polymodels-y0}. The inset shows the right end, magnified.}}
\centering
\includegraphics[width=0.88 \columnwidth]{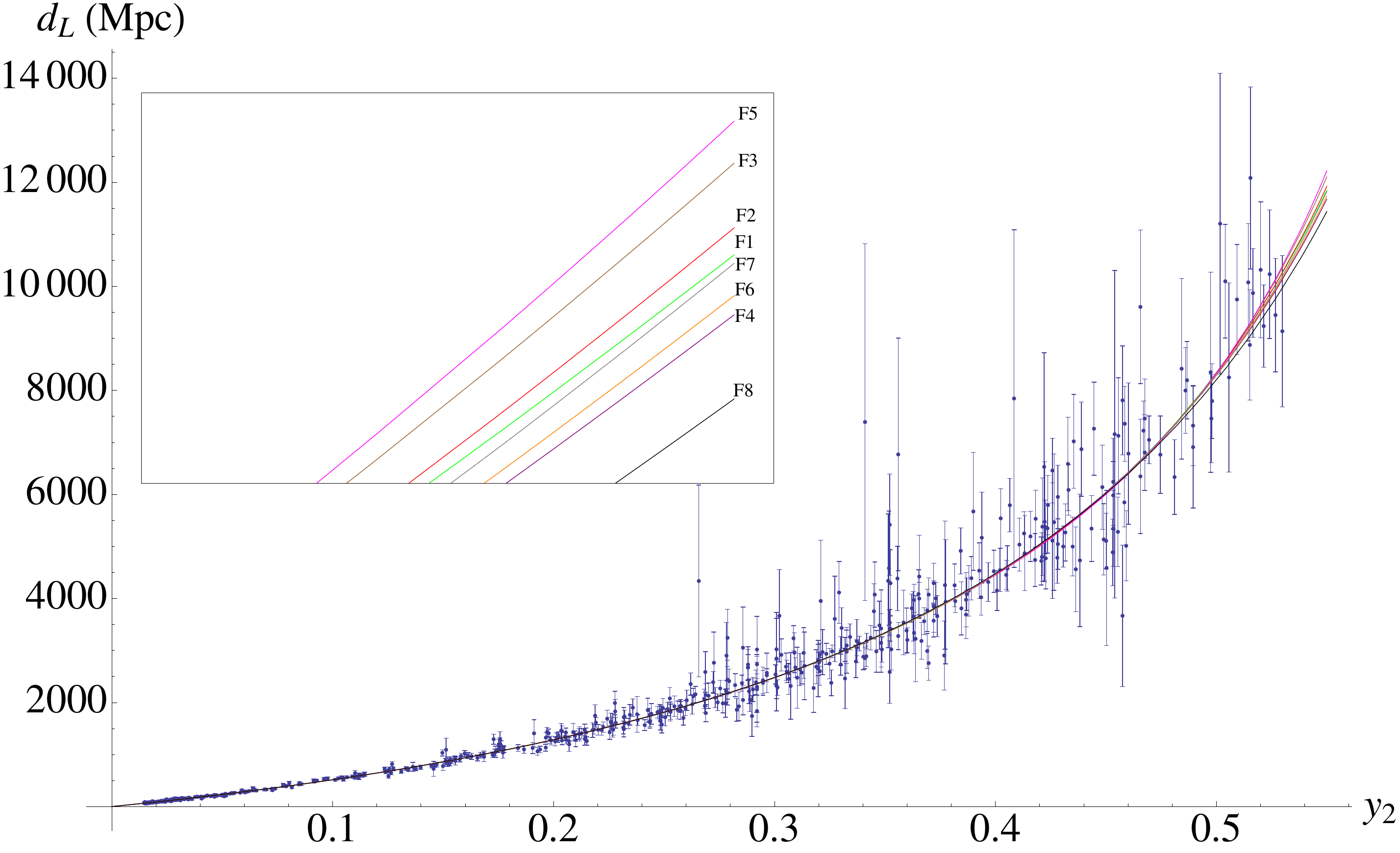}
\label{fig:bestfits-y2} 
\end{figure}

\begin{figure}[h!]
\caption{\textsf{Union 2.1 data (in terms of luminosity distance and the redshift variable $y_{4}$; and the best fits for each function family listed in Table \ref{families}. The one-sigma confidence-levels are not shown in order not to clutter up the figure; they are similar to those in Fig.\ref{fig:polymodels-y0}. The inset shows the right end, magnified.}}
\centering
\includegraphics[width=0.88 \columnwidth]{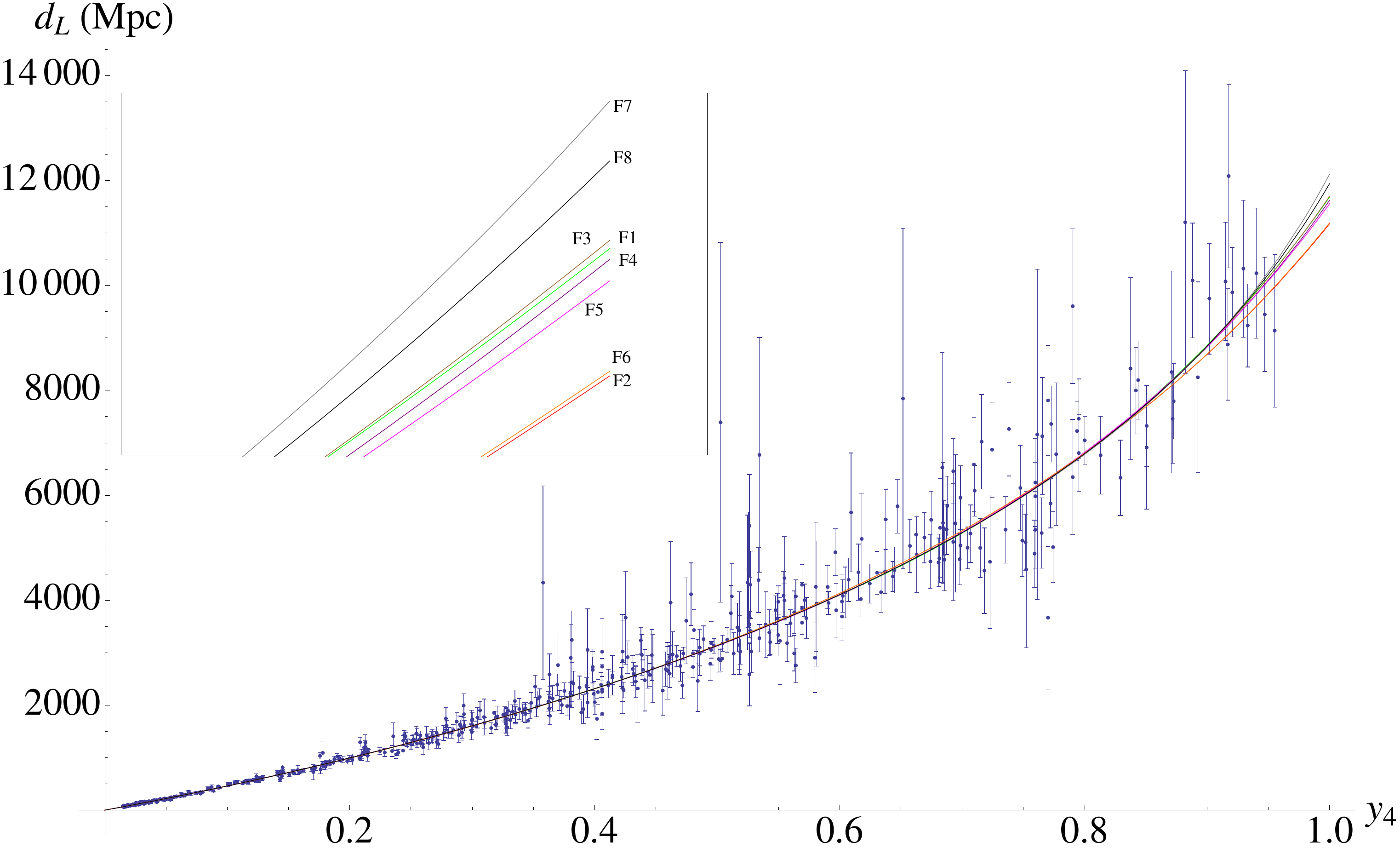}
\label{fig:bestfits-y4} 
\end{figure}

\begin{figure}[h!]
\caption{\textsf{Union 2.1 data (in terms of luminosity distance and the redshift variable $y_{5}$; and the best fits for each function family listed in Table \ref{families}. The one-sigma confidence-levels are not shown in order not to clutter up the figure; they are similar to those in Fig.\ref{fig:polymodels-y0}.  The inset shows the right end, magnified.}}
\centering
\includegraphics[width=0.88 \columnwidth]{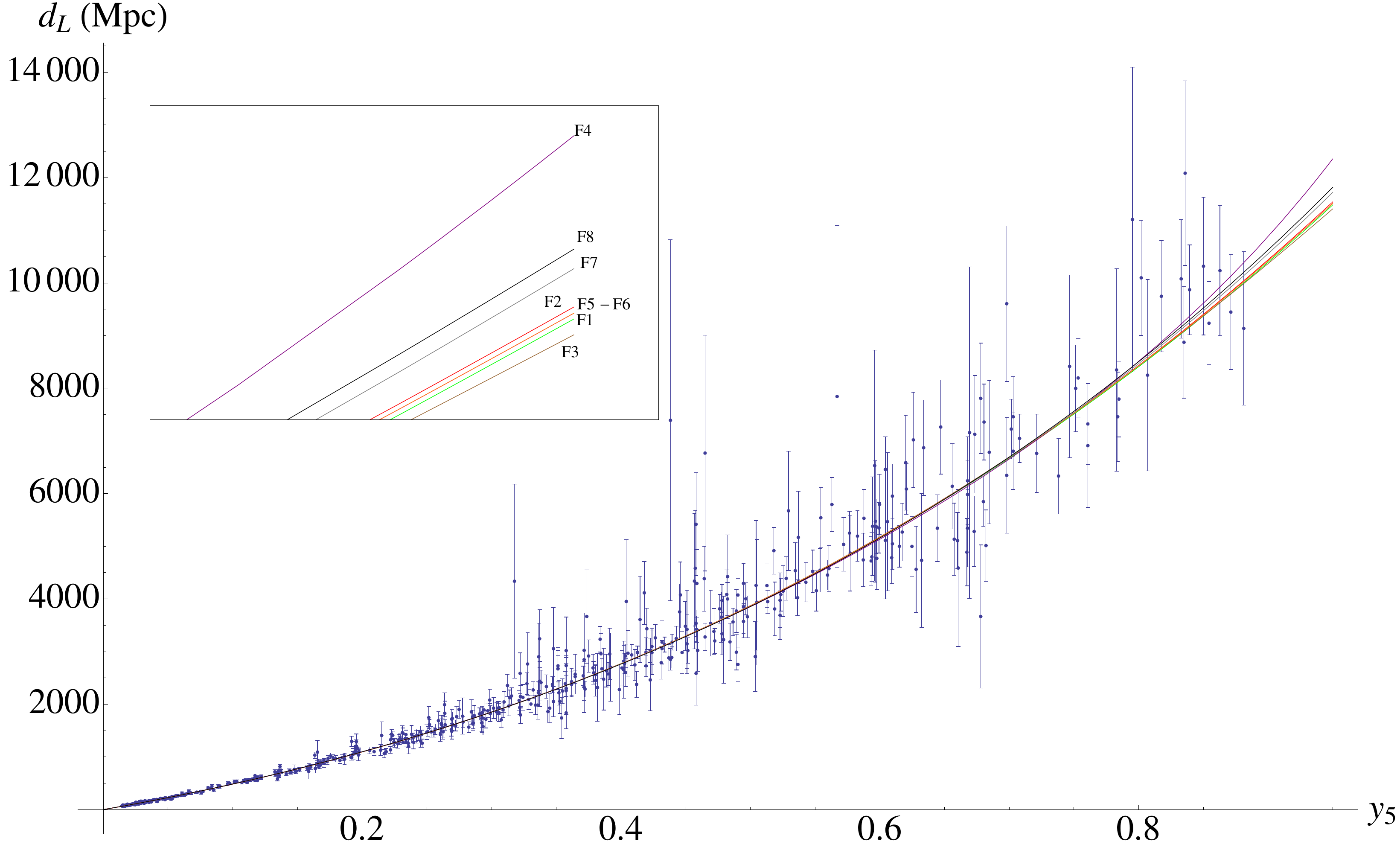}
\label{fig:bestfits-y5} 
\end{figure}

\begin{figure}[h!]
\caption{\textsf{Union 2.1 data (in terms of luminosity distance and the redshift variable $y_{6}=u=1+z$; and the best fits for each function family listed in Table \ref{families}. The one-sigma confidence-levels are not shown in order not to clutter up the figure; they are similar to those in Fig.\ref{fig:polymodels-y0}.  The inset shows the right end, magnified.}}
\centering
\includegraphics[width=0.99 \columnwidth]{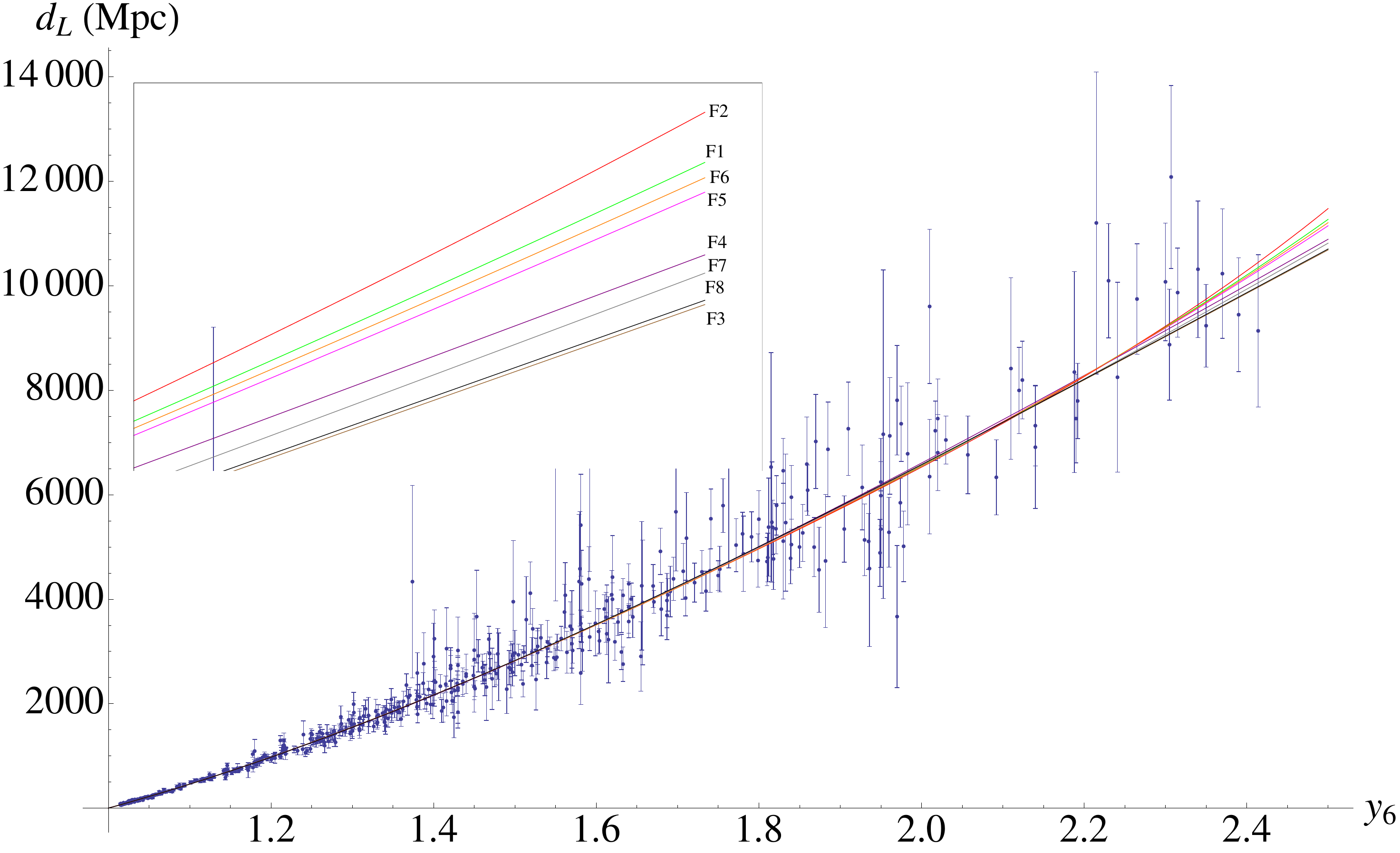}
\label{fig:bestfits-y6} 
\end{figure}

The full sets of best fits for each family and redshift variable are indicated in Table \ref{families}.
\begin{table} [h!]
\caption{\textsf{The best fits for the Union 2.1 SNe Ia data. Each cell displays the internal label of the best-fitting member of the family for that row and the $\chi^2_\nu \equiv \chi^2$/d.o.f value associated with the redshift variable for that column. Note that the $\chi^2_\nu$ value for the $\Lambda$CDM model is 0.9340, (with $y_{0}=z$ as the independent variable)}
}
\centering
\resizebox{14cm}{!}{
\begin{tabular}{| c | c | c | c | c | c | c |}
\hline
\backslashbox{family}{variable}& $y_{0}=z$ & $y_{1}$ & $y_{2}$ & $y_{4}$ & $y_{5}$ & $y_{6}=u$  \\
\hline
F1 & 4; 0.9355 &  5; 0.9371 & 6; 0.9384 & 5; 0.9366 & 3; 0.9354 & 4; 0.9355 \\
\hline
F2 & 4; 0.9353 & 3; 0.9353 & 5; 0.9367 & 3; 0.9360 & 3; 0.9353 & 4; 0.9353 \\
\hline
F3 & 5; 0.9370 &  4; 0.9367 & 6; 0.9381 & 4; 0.9353 & 3; 0.9365 & 3; 0.9361 \\
\hline
F4 & 2; 0.9357 &  2; 0.9340 & 4; 0.9365 & 4; 0.9358 & 4; 0.9354 & 2; 0.9357 \\
\hline
F5 & 4; 0.9357 & 5;  0.9372 & 6; 0.9381 & 5; 0.9368 & 3; 0.9355 & 4; 0.9357 \\
\hline
F6 & 4; 0.9356 & 3; 0.9351 & 3; 0.9358 & 3; 0.9359 & 3; 0.9355 & 4; 0.9356 \\
\hline
F7 & 2;1; 0.9350 & 2;1; 0.9344 & 2;1; 0.9343 & 3;2; 0.9384 & 2;1; 0.9346 & 2;1; 0.9350 \\
\hline
F8 & 1;2; 0.9364 & 2;1; 0.9350 & 1;1; 0.9339 & 2;2; 0.9370 & 2;1; 0.9348 & 1;2; 0.9364 \\
\hline
\end{tabular}
}
\label{families}
\end{table}
Fig.\ref{fig:bestfits-y0} shows the best-fitting members from each family, for the variable $z \equiv y_{0}$. The one-sigma confidence-levels are not shown in order not to clutter up the figure; they are similar to those in Fig.\ref{fig:polymodels-y0}, unless stated otherwise. Figures \ref{fig:bestfits-y1}-\ref{fig:bestfits-y6} are similar to Fig.\ref{fig:bestfits-y0}, but for the other redshift variables.


We see from Table \ref{families} that none of the fits  is significantly better than the $\Lambda$CDM model. In fact, only one (F8, $y_{2}$)  has a lower and one (F4, $y_{1}$) an equal $\chi^2_\nu$ value; they range between  0.9339 and 0.9384, compared to $\Lambda$CDM's 0.9340 (with $y_{0}=z$ as the independent variable). But, being the best fits in their respective families, they are not significantly worse either, so we will use all of them for the time being. 

\section{Determination of the expansion history of the universe}

\subsection{More on the procedure}

As declared in the introduction, what we would ideally like to do is to find $t(z)$ by performing the integration (\ref{lookback}) for a given $d_{L}(z)$ function, then find $z(t)$ by inverting that result, hence find $a(t)$ by eq.(\ref{scalefactorredshift}) [Of course, the same procedure can also be applied in terms of alternative variables $y_{i}$].
 Unfortunately, for all but the simplest functions, neither the intergration, nor the subsequent inversion can be analytically performed.  Numerical integration and inversion can provide numerical $a(t)$ functions, but numerical differentiation of these is problematic, hence not very useful.

However, for a given $d_{L}(z)$, the time-derivatives of $a(t)$ can be calculated {\it analytically} in terms of the redshift variable(s):
\begin{equation}
\dot{a} = \frac{da}{dt} = \frac{da}{dz} \frac{dz}{dt}
\label{eq:a-dot-of-z}
\end{equation}
i.e. as the $z$-derivative of (\ref{scalefactorredshift}) divided by  the integrand in (\ref{lookback}). Similarly,
\begin{equation}
\ddot{a} = \frac{d}{dt} \dot{a} = \frac{d\dot{a}}{dz} \frac{dz}{dt}.
\label{eq:a-ddot-of-z}
\end{equation}
Furthermore, one can plot (but not write) $\dot{a}$ and $\ddot{a}$ as functions of time, since $z(t)$ can be evaluated numerically by integration, which is a robust procedure.

We also noted in the Introduction that the SNe Ia data are meaningful for any value of curvature; hence the expression $dz/dt$, coming from eq.(\ref{lookback}) and used in eqs.(\ref{eq:a-dot-of-z}) and (\ref{eq:a-ddot-of-z}) contains the curvature $\kappa$. We choose \mbox{$\kappa_{0} = (10000 \rm{Mpc})^{-2} \sim (2c/H_0)^{-2}$} as the upper limit to positive curvature; $10000$ Mpc is also approximately the maximum $d_L(z)$ value in the Union2.1 data. We perform the procedure for $\kappa$ values between $-\kappa_{0}$ and $+\kappa_{0}$, in $0.2\kappa_{0}$ increments, i.e. we take $\kappa = k' \kappa_{0}$, with $-1 \leq k' \leq +1$.

Incidentally, the current value of the Hubble parameter can be extracted from $d_{L}(z)$ as
\begin{equation}
H_{0} = \left. \frac{\dot{a}}{a} \right|_{z=0}  = \frac{c}{\left. \frac{d}{dz}\left[ \frac{d_L(z)}{1+z}\right]\right|_{z=0}},
\label{eq:Hubble_param}
\end{equation}
and similar expressions can be written in terms of the other redshift variables. We find $H_{0} = 70.55 \pm 0.62 \frac{\rm km/s}{\rm Mpc}$, using all best fits in Table \ref{families}.

\subsection{Determination of time derivatives of the scale function}

\begin{figure}[h!]
\caption{\textsf{The $a(t)$ functions, computed by numerical integration of eq.(\ref{lookback}) and subsequent numerical inversion to find $z(t)$ (or similar analysis with one of the redshift variables $y_{i}$), using the fits found for $d_{L}(y_{i})$ to the Union 2.1 data. The columns represents analyses with $y_0, y_1, y_2, y_4, y_5$ and $u$,  respectively, and rows are for candidate families F1-F8. In each plot, the horizontal axis is $t$, ticked at 1 Gyr intervals, and the vertical axis is $a(t)$, normalized to 1 at the present epoch shown at the right end of each plot.  For curvature, the same color-coding is used as in figure \ref{fig:adot(t)graph}, but the curves for different curvatures practically overlap in this figure. }}
\centering
\includegraphics[width=0.99 \columnwidth]{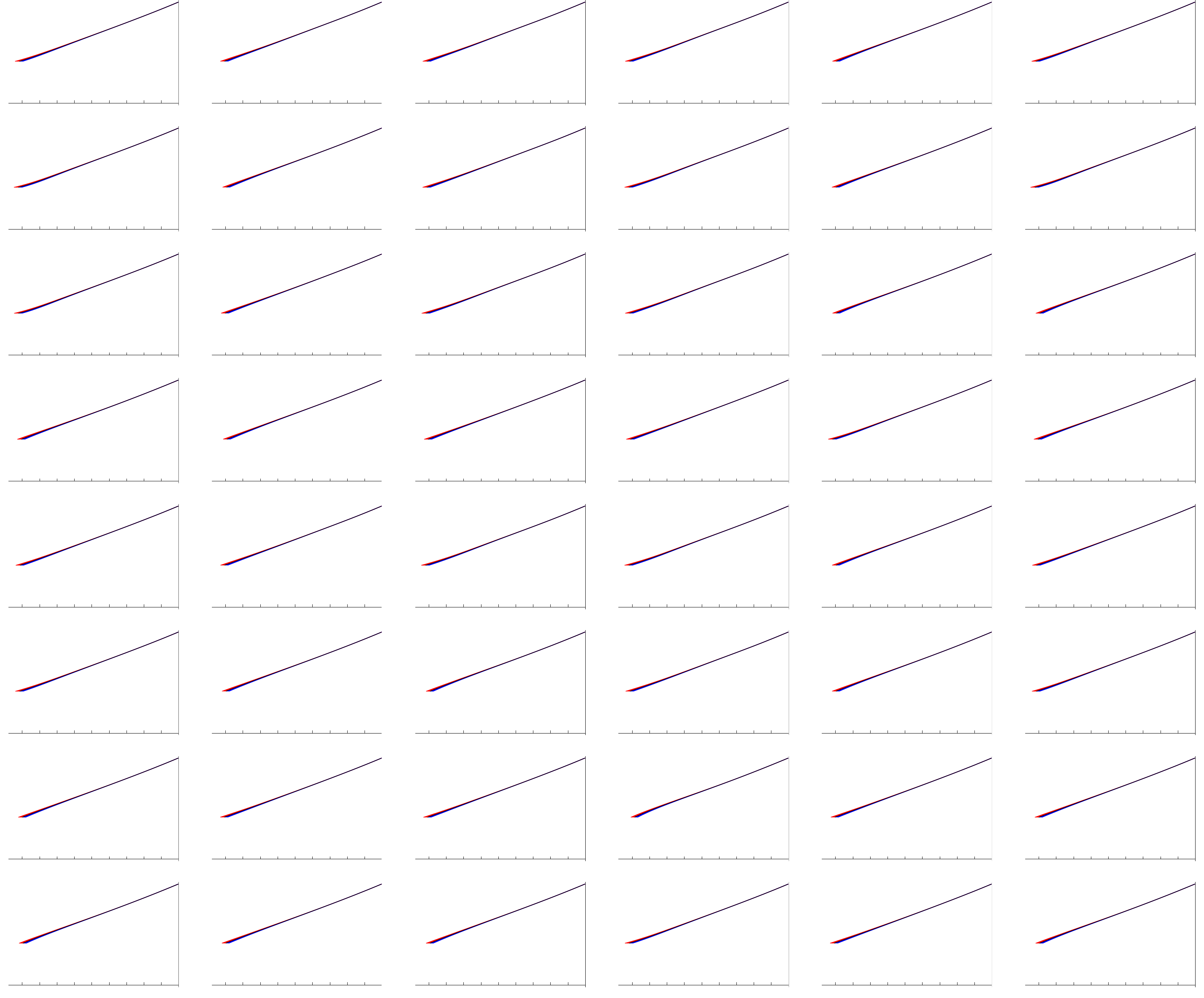}
\label{fig:a(t)graph} 
\end{figure}

\begin{figure}[h!]
\caption{\textsf{The $\dot{a}(t)$ functions, computed analytically by eq.(\ref{eq:a-dot-of-z}) or its analogs for the different redshift variables $y_{i}$. As in the previous figure, the columns represents analyses with $y_0, y_1, y_2, y_4, y_5$ and $u$, respectively, and rows are for candidate families F1-F8 for $d_{L}(y_{i})$, fitted to the Union 2.1 data. Blue, red and black curves represent open, closed and flat spaces respectively, with the range $-\kappa_{0} \leq \kappa \leq \kappa_{0}$. For comparison purposes, the horizontal axes are transformed to be $z=y_{0}$, ticked with  intervals $\Delta z=0.2$, and $\dot{a}(t)$ is normalized to 1 in ($H_{0}$$a_{0}$)  units.}}
\centering
\includegraphics[width=0.99 \columnwidth]{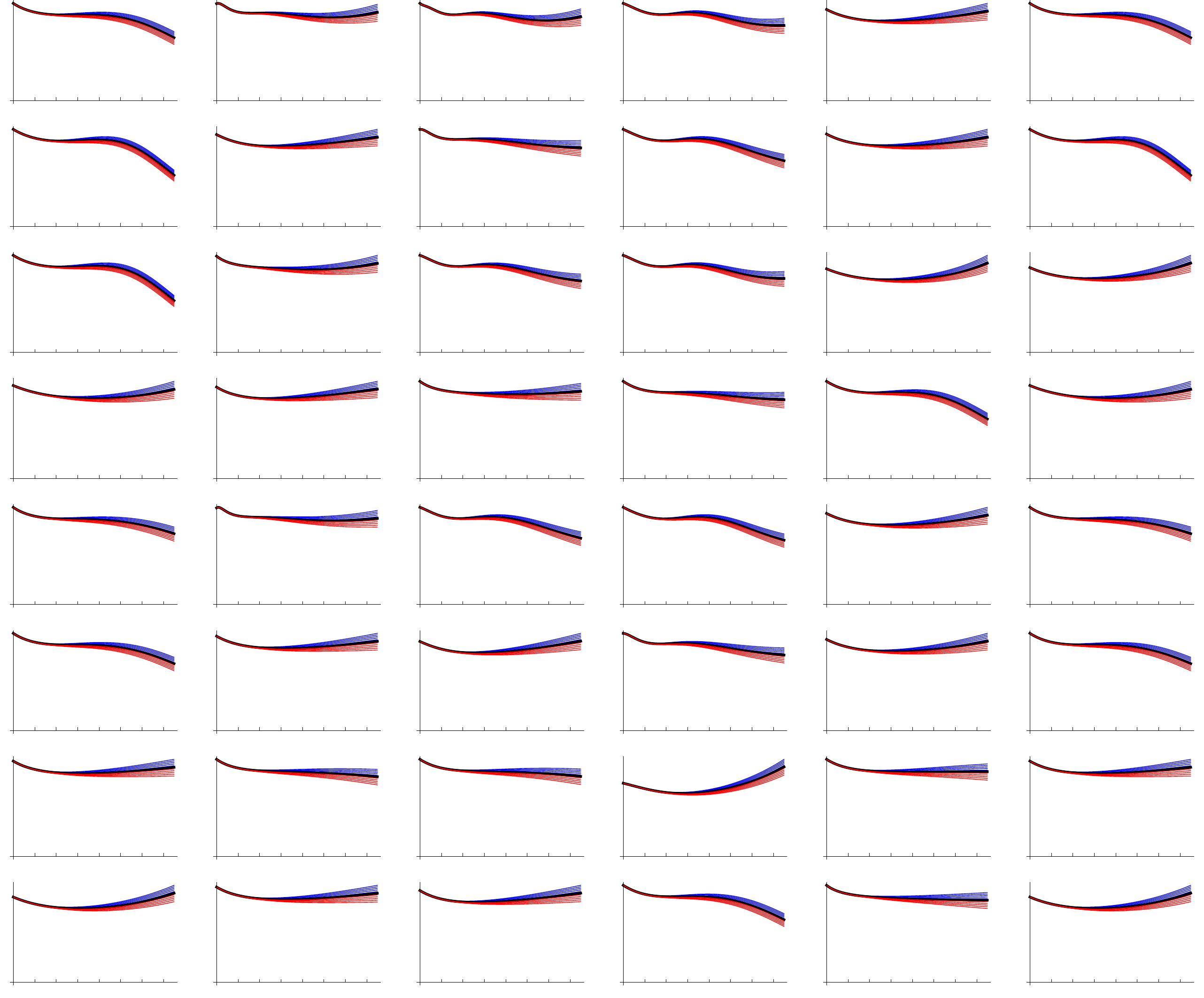}
\label{fig:adot(t)graph} 
\end{figure}

\begin{figure}[h!]
\caption{\textsf{All $\ddot{a}(t)$ functions, computed analytically by eq.(\ref{eq:a-ddot-of-z}) or its analogs for the different redshift variables $y_{i}$. As in the previous figure, the columns represents analyses with $y_0, y_1, y_2, y_4, y_5$ and $u$, respectively, and rows are for candidate families F1-F8 for $d_{L}(y_{i})$, fitted to the Union 2.1 data, with same color coding. For comparison purposes, the horizontal axes are transformed to be $z=y_{0}$, ticked with  intervals $\Delta z=0.2$, and the vertical axes are in arbitrary units.}}
\centering
\includegraphics[width=1 \columnwidth]{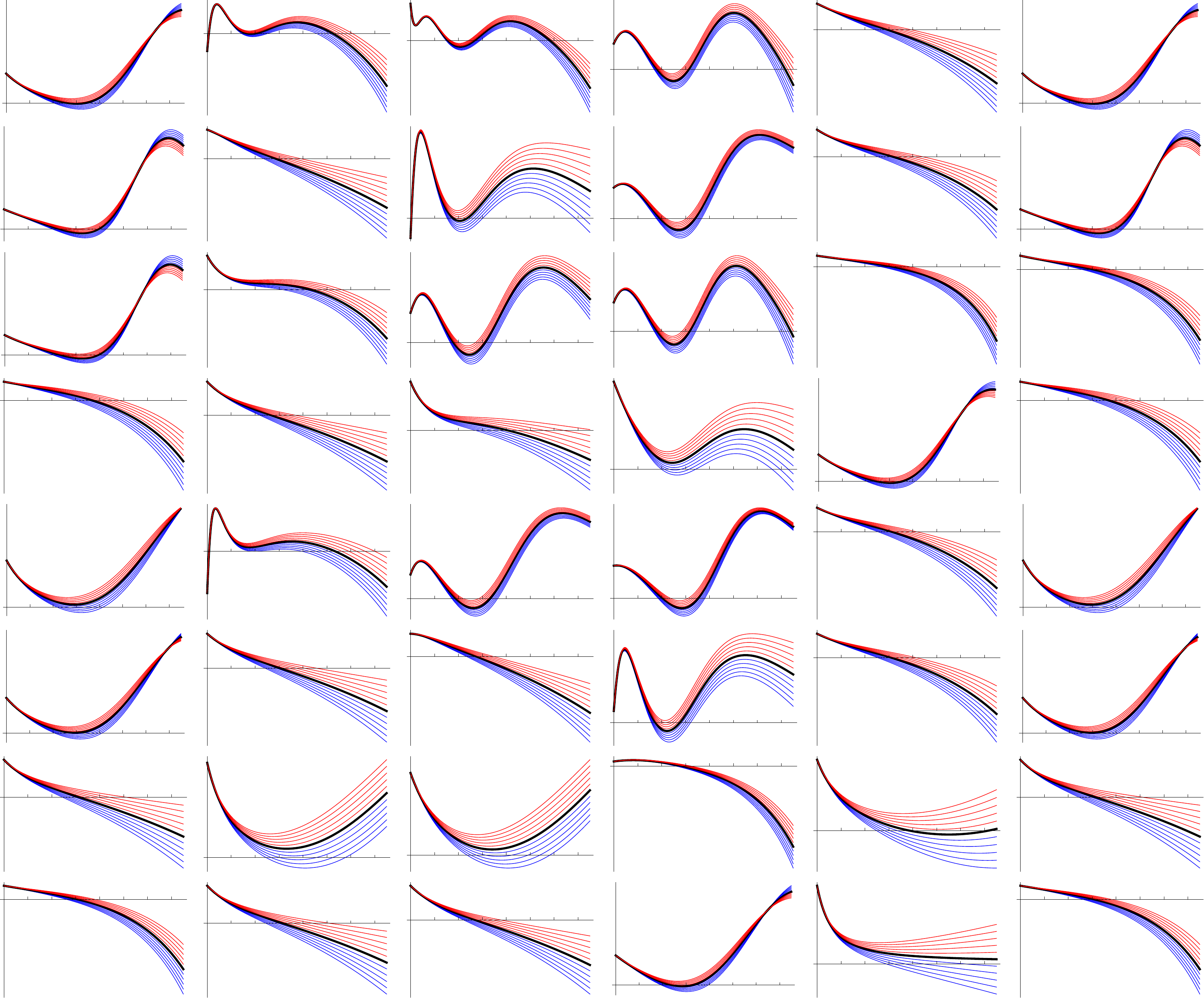}
\label{fig:grid-a-ddot-supernova} 
\end{figure}

The use of 6 different redshift variables and 8 function families for approximating $d_L(z)$ gives 48 possible representations of the  expansion history of the universe. 
When we plot the possibilities for the scale function $a(t)$ in a \mbox{6 $\times$ 8} grid (Fig.\ref{fig:a(t)graph}), the  repesentations are visually virtually identical, and not very informative.
The $\dot{a}(z)$ plots shown in Fig.\ref{fig:adot(t)graph} are a bit more meaningful visually,
but the really visually informative plots are those of acceleration, $\ddot{a}$, where sign changes are more apparent (Figure \ref{fig:grid-a-ddot-supernova}).
All these plots were made for 11 different spatial curvature values chosen as described previously.

But, problems with the $\ddot{a}(t)$ plots  strike the eye at first look.  Many of the plots show unnatural-looking fluctuations; especially divergences at high $y_i$ due to differentiating the $d_L$ twice: Powers of $z$ with coefficients arranged to mostly cancel each other do not do so when those coefficients are changed due to differentiation. As a result, some models would seem to suggest negative acceleration today, while some others show positive acceleration in the past. So one can see that using the current SNe Ia data in a model-independent way, one cannot tell the value, or even the existence of the transition redshift; as also discussed in \cite[Appendix A]{ln(1+z)}, and references therein, conclusions can heavily depend on the choice of parametrization (see however, \cite{transitionVar} for a dissenting view).

In terms of showing least amount of unnatural fluctuations, the best variable seems to be the newly suggested redshift variable $y_5$, (the fifth column) the variables $y_{2}$ and $y_{4}$ showing significantly more fluctuations. One might speculate if this has to do with these variables being related to $a/a_{0}$ via trigonometric relations. Among the functional forms for $d_{L}(y_{i})$, least amount of unnatural fluctuations in $\ddot{a}(z)$ are featured by the families featuring the Pade approximant (the two lowest rows).

This last statement gives a hint to the source of the problem: All but the last two families in Table \ref{introducingfamilies} have a polynomial as a factor, and polynomial fits to a set of data quickly diverge away from the visual data pattern just outside the interval with the data; due to the dominance of the highest powers at large magnitudes of the independent variable. Absence of this behavior is the main advantage of the Pade approximation over truncated McLaurin series (e.g.\cite{pade}).

\begin{figure}[t!]
\caption{\textsf{The Union 2.1 supernova data, shown in blue, the GRB data \cite{GRB}, shown in red, and the Quasar \cite{quasars} data shown in black, together.}}
\centering
\includegraphics[width=0.99 \columnwidth]{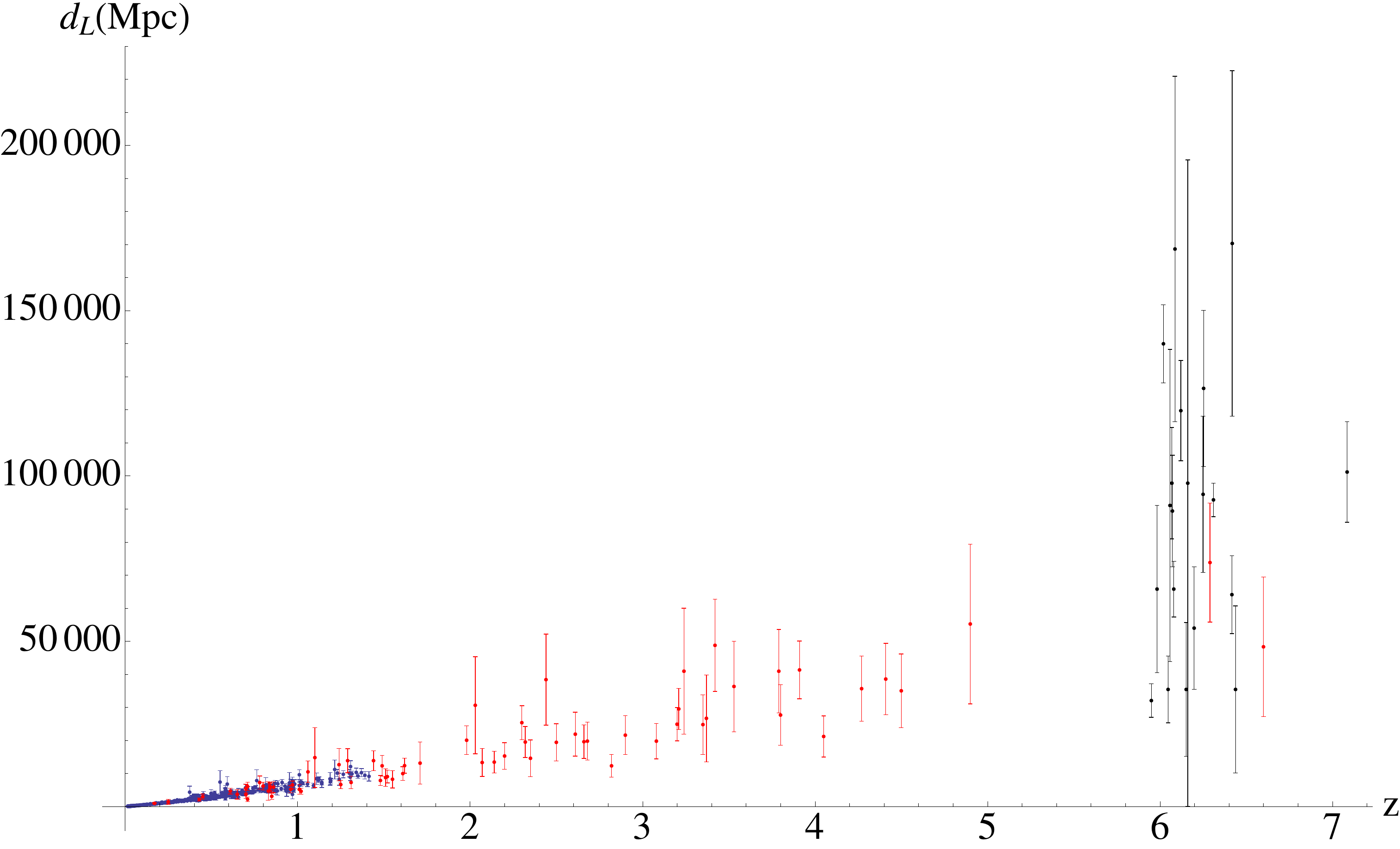}
\label{fig:extdata} 
\end{figure}

This divergence can be tamed by using data in a wider interval, if available. In our case, data extending to $z\sim6$ {\it are} available as Gamma-Ray Burst \cite{GRB} and Quasar \cite{quasars} data (Fig.\ref{fig:extdata}),
but they actually are not suitable for model-independent analysis: They are called {\it standardisable} candles, and are not really {\it standard} candles, since the interpretation of the measurements depends on the assumed model of cosmology. They also have large errors. Nevertheless, they can serve the purpose of taming the divergence of polynomials, while their large errors will give them little weight in the fits, hence they will not contaminate the analysis \textit{very much} with physical model assumptions.

\begin{figure}[h!]
\caption{\textsf{The $\ddot{a}(t)$ functions, computed analytically by eq.(\ref{eq:a-ddot-of-z}) or its analogs for the different redshift variables $y_{i}$. The rows, columns and color coding have the same meaning as in  figures \ref{fig:a(t)graph}-\ref{fig:grid-a-ddot-supernova}, fitted to the Union 2.1 and GRB data. For comparison purposes, the horizontal axes are transformed to be $z=y_{0}$, ticked with  intervals $\Delta z=0.2$,and the vertical axes are in arbitrary units.}}
\centering
\includegraphics[width=0.99 \columnwidth]{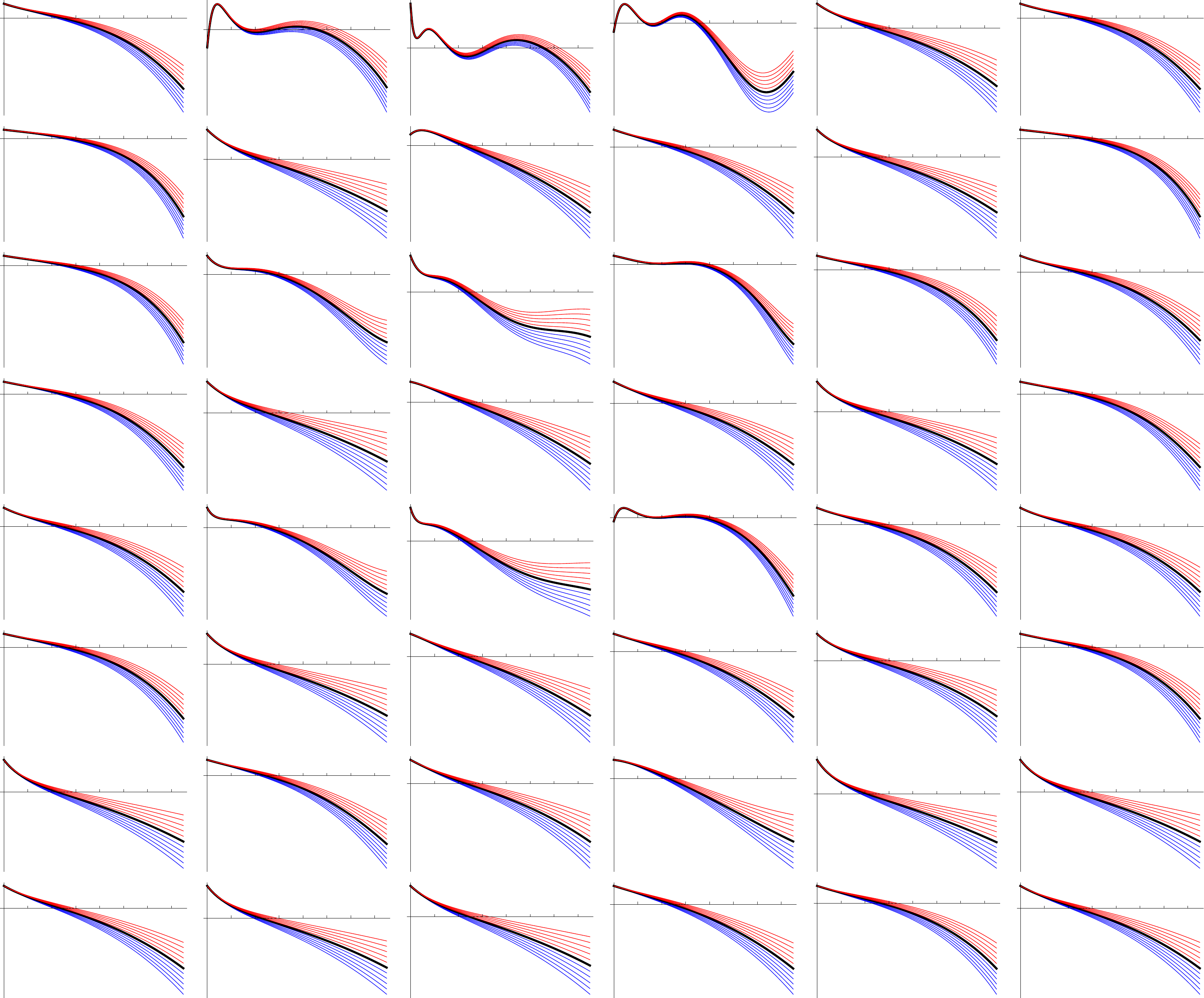}
\label{fig:grid-a-ddot-supernova-grb} 
\end{figure}

Unfortunately, the errors in the Quasar data are too big: Incorporating them in our data set increases the $\chi^2_\nu$ value to unacceptably large values, around 1.2. They are also concentrated near $z \sim 6$, whereas the GRB data are spread over the whole range $0 < z < 7$. Therefore we choose to incorporate only the GRB data consisting of 69 GRB's. Their  addition to the 580 SNe Ia data modifies the plots of $\ddot{a}$ to give Fig.\ref{fig:grid-a-ddot-supernova-grb};
\begin{table} [h!]
\caption{\textsf{The best fits for the Union 2.1 SNe Ia + GRB data. Each cell displays the internal label of the best-fitting member of the family for that row and the $\chi^2_\nu \equiv \chi^2$/d.o.f value associated with the redshift variable for that column.}}
\centering
\resizebox{14cm}{!}{
\begin{tabular}{| c | c | c | c | c | c | c |}
\hline
\backslashbox{family}{variable}& $y_{0} = z$ & $y_{1}$ & $y_{2}$ & $y_{4}$ & $y_{5}$ & $y_{6}=u$  \\
\hline
F1 & 5; 0.9540 & 5; 0.9591 & 6; 0.9616 & 7; 0.9594 & 3; 0.9529 & 5; 0.9540 \\
\hline
F2 & 5; 0.9540 & 3; 0.9530 & 3; 0.9531 & 3; 0.9527 & 4; 0.9542 & 5; 0.9540 \\
\hline
F3 & 6; 0.9550 & 6; 0.9569 & 6; 0.9588 & 6; 0.9560 & 4; 0.9536 & 5; 0.9543 \\
\hline
F4 & 5; 0.9543 & 2; 0.9526 & 4; 0.9539 & 4; 0.9539 & 3; 0.9532 & 5; 0.9543 \\
\hline
F5 & 4; 0.9530 & 6; 0.9567 & 6; 0.9583 & 7; 0.9564 & 5; 0.9543 & 4; 0.9530 \\
\hline
F6 & 4; 0.9533 & 3; 0.9532 & 3; 0.9524 & 3; 0.9526 & 4; 0.9545 & 4; 0.9533 \\
\hline
F7 & 2;1; 0.9534 & 2;3; 0.9545 & 1;2; 0.9526 & 3;2; 0.9551 & 2;1; 0.9539 & 2;1; 0.9534 \\
\hline
F8 & 2;2; 0.9554 & 2;1; 0.9534 & 1;1; 0.9525 & 1;2; 0.9540 & 1;2; 0.9554 & 2;2; 0.9554 \\
\hline
\end{tabular}
}
\label{families2}
\end{table} 
and the $\chi^2_\nu$ values in Table \ref{families2}.
It can be seen that the unnatural fluctuations have disappeared from most of the plots. Moreover, deceleration in the remote past can be seen in all the plots, unlike in Figure \ref{fig:grid-a-ddot-supernova}.  As in that figure, the fifth column and seventh row are free from unnatural fluctuations, but so are the the first and sixth columns; and the second, fourth, sixth and eighth rows. Interestingly, presence of the factor $u(y)=1+z$ improves the fit. In fact, in almost every cell of rows two, four and six of Table \ref{families}, the $\chi^2_\nu$ value is smaller than the cell just above it. This motivated the addition of $u=z+1$ as an extra redshift variable to the analysis (and of family F8). It turned out that this trivial-looking shift in the independent variable can serve as an illustrative example of the parametrization-dependence of conclusions refered to above, in the discussion of Fig. \ref{fig:grid-a-ddot-supernova}: The best-fits for the third family are quite different! A close inspection of Table \ref{families} reveals that after the variable shift, the three-parameter member of the family is the best fit, whereas it was the five-parameter one before. The results for the other families are not much affected, and the inclusion of the GRB data makes the difference much less drastic.

\begin{figure}[h!]
\caption{\textsf{The average of the fifteen ``natural'' $\ddot{a}(t)$ functions from Fig.\ref{fig:grid-a-ddot-supernova-grb}, namely those in columns one, five and six, and rows two, four, six, seven and eight, extended to negative $z$, i.e. to the future. The color coding has the same meaning as in  figures \ref{fig:a(t)graph}-\ref{fig:grid-a-ddot-supernova}. The horizontal axis shows $z=y_{0}$, ticked with  intervals $\Delta z=0.1$, and the vertical axis is in arbitrary units.}}
\centering
\includegraphics[width=0.99 \columnwidth]{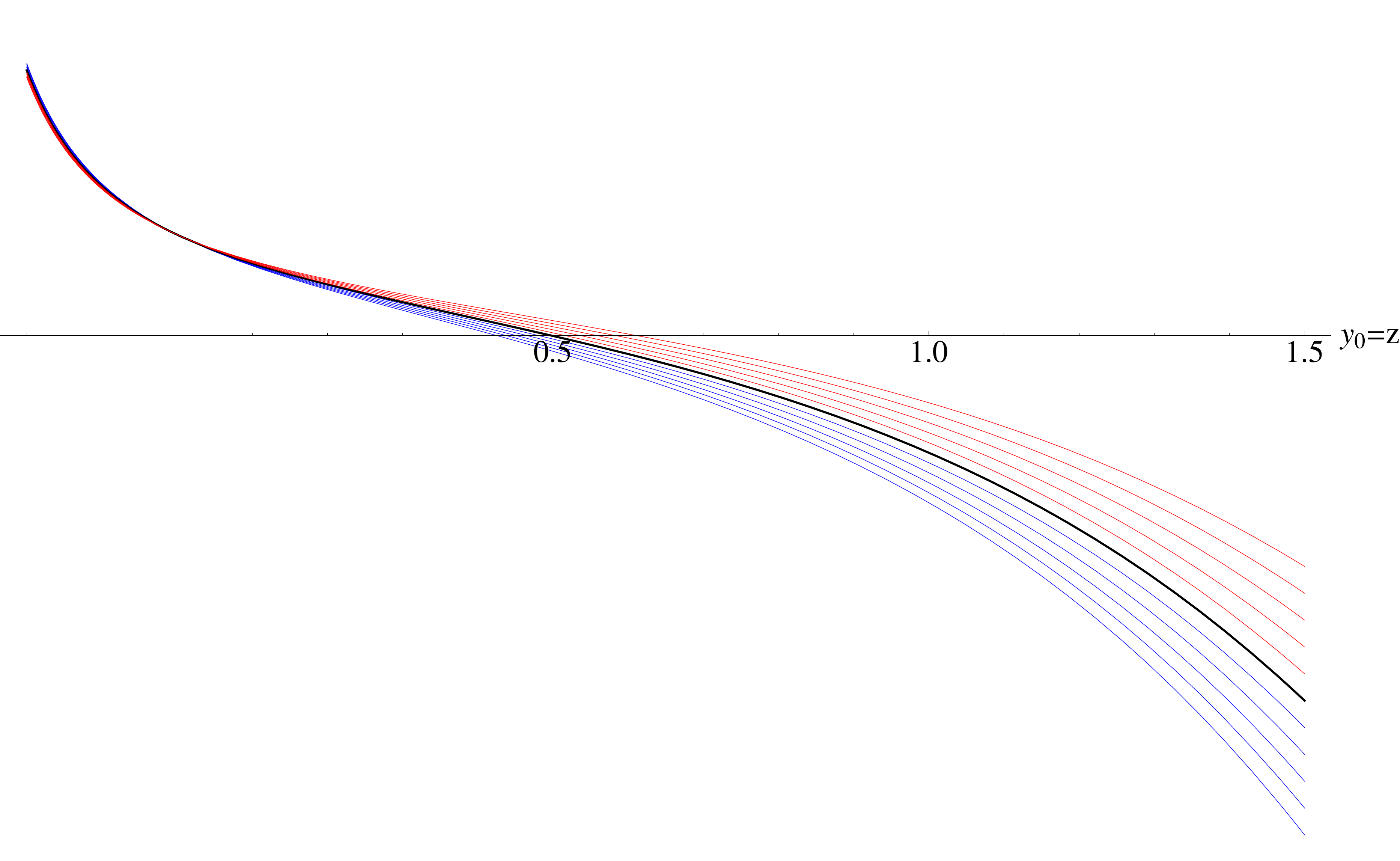}
\label{fig:supernova-grb-avg-acc} 
\end{figure}

So, from here on in this work, we will use only the ``more natural'' fits, i.e. the fifteen fits belonging to columns one, five and six, and rows two, four, six, seven and eight of Fig.\ref{fig:grid-a-ddot-supernova-grb}. We show the average of these fifteen graphs in Fig.\ref{fig:supernova-grb-avg-acc}. For the confidence interval, we take at each point the maximum of the confidence intervals of the fifteen curves. We do not apply the usual combination of errors technique, since these  curves do not correspond to different ``measurements'', they are different representations of the same one. 

Consideration of these gives for the redshift of transition from deceleration in the past to current acceleration as
\begin{equation}
z_{\rm t, flat} = 0.50^{+0.08}_{-0.10}
\label{eq:z-trans-flat}
\end{equation}
for the case the universe is flat, and
\begin{equation}
z_{\rm t, +} = 0.61^{+0.12}_{-0.13},   \;\;\  z_{\rm t, -} = 0.42^{+0.08}_{-0.08}
\label{eq:z-trans-curved}
\end{equation}
for the most positively and most negatively curved universes.

In that figure, we also show an extension towards negative $z$, i.e the future --after all, the curves represent analytical functions--, predicting that the acceleration will continue to increase.

\subsection{Is the universe accelerating {\it now}?} \label{ssect:accelnow}

In figures \ref{fig:grid-a-ddot-supernova} and \ref{fig:grid-a-ddot-supernova-grb}, we see that as $z\rightarrow 0$, $\ddot{a}$ approaches the same value  for all curvatures. This may seem surprising, since there is an impression that the supernova data cannot unequivocally show that the expansion of the universe is accelerating (e.g. \cite[Figure 1]{mortsellclarkson}) and one needs complementary and independent data on the curvature of the universe (e.g. from the power spectrum of the CMB temperature anisotropies). 
To address this question, let us investigate the analytical relationship between $\ddot{A}_{\rm{now}}$ and $d_L(z)$, where $A$ is a shorthand for $a/a_0$. $\ddot{A}$ can be evaluated in terms of redshift, using (\ref{scalefactorredshift}):
\begin{equation}
\ddot{A}=\frac{2}{(1+z)^3}\dot{z}^2-\frac{1}{(1+z)^2}\ddot{z},
\end{equation} 
where $\dot{z}$ can be plugged in from (\ref{lookback})
\begin{equation}
\dot{z}=\frac{-c(1+z)\sqrt{1-\kappa\frac{d_L^{2}}{(1+z)^2}}}{\frac{d}{dz}\left[\frac{d_L}{1+z}\right]}=f(z)\sqrt{1-\kappa D^{2}(z)},
\end{equation}
where $D(z)=d_L(z)/(z+1)$, sometimes called as ``photon count distance" \cite{cosmography}, and $f(z)$ is the part outside the square root. Now, acceleration takes the form
\begin{equation}
\ddot{A}=\frac{2}{(1+z)^3}\left(f(z)\sqrt{1-\kappa D^{2}(z)}\right)^2-\frac{1}{(1+z)^2}\frac{d}{dt}\left(f(z)\sqrt{1-\kappa D^{2}(z)}\right).
\end{equation}
For $z\rightarrow 0$ it is easy to show that $D(z)\rightarrow 0$. Recalling $\dot{(D^{2})} = 2 D \dot{D}$, we have
\begin{equation}
\ddot{A}_{\rm{now}}=\left(2f^2-\dot{f}\right)_{\rm{now}} =\left(2f^2-f'\dot{z}\right)_{\rm{now}}
\end{equation}
where $(')$ denotes derivative with respect to $z$. Putting $\dot{z}$ in once more,
\begin{equation}
\ddot{A}_{\rm{now}}=\left(2f^2-f'f\right)_{\rm{now}}.
\end{equation}
Finally replacing $f(z)$ and rearranging terms, 
\begin{equation}
\ddot{A}_{\rm{now}}=\left(\frac{c^2}{d_L'^3}\left(d_L''-d_L'\right)\right)_{\rm{now}}.
\end{equation}
Thus we have shown that the present acceleration is independent of spatial curvature for given $d_L(z)$,  and can be evaluated from data without assuming any theory of gravity. This conclusion, first remarked in \cite{spacecurvature} does not seem to have gotten sufficient interest from the scientific community.  

\section{Inferences in GR about the content of the universe}

\begin{figure}[h!]
\caption{\textsf{The density of the universe as function of $u = z+1$, as calculated using the redshift variable $y5$ and fit family F7, assuming Einstein gravity. The color-coding is the same as used in figure \ref{fig:adot(t)graph}. Note the intersection around $z \approx 1.5$.}}
\centering
\includegraphics[width=0.99 \columnwidth]{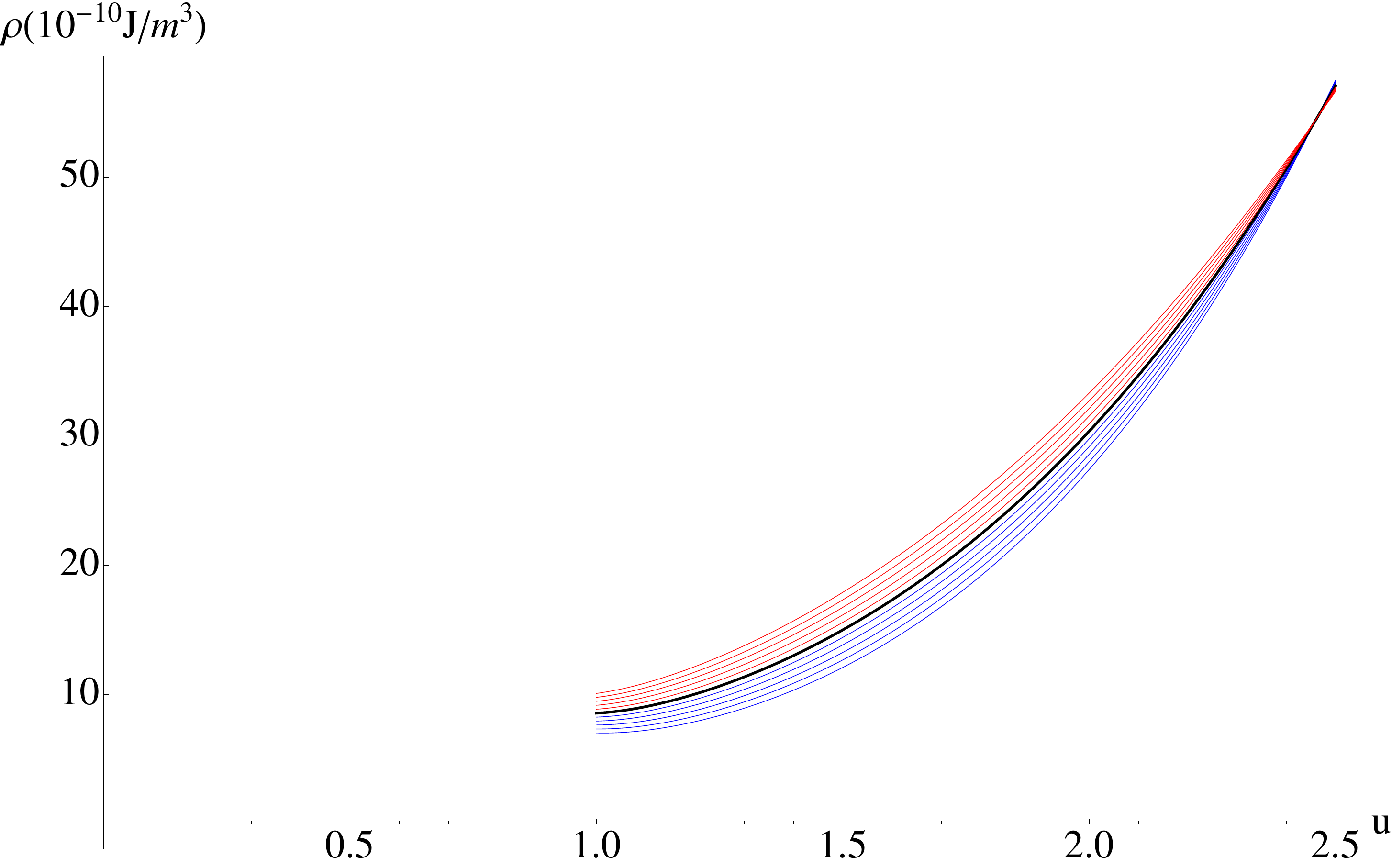}
\label{fig:rho_y5_f7} 
\end{figure}

Up to this point we did not specify any theory of gravitation in our research. However it is appropriate now to assume General Relativity so that we can comment on the matter-energy content of the universe. First Einstein's equation in FRW framework gives,

\begin{equation}
H^{2}+\frac{k c^2}{a^2}=\frac{8\pi G}{3c^2}\rho \label{firsteinstein}
\end{equation}
hence it is straightforward to construct $\rho(z)$ using (\ref{eq:a-dot-of-z}). Using $u = z+1$ as independent variable for future convenience, we plot an example for the evolution of matter-energy density, for different spatial curvatures in Fig.(\ref{fig:rho_y5_f7}). This particular figure was created using redshift variable $y5$ and fit family F7. 

An interesting feature manifests itself in this figure; there is an intersection point about $z\sim1.5$ for curves with different spatial curvatures. To check the existence of this point analytically we plug (\ref{lookback}) in (\ref{firsteinstein}) and get
\begin{equation}
\rho(z)=\frac{3c^2}{8\pi G}\left(c^2(1-\kappa D^2)\left(\frac{dD}{dz}\right)^{-2}+c^2\kappa(z+1)^2 \label{rho_kappa_z}\right)
\end{equation}
where $D(z)$ was defined in Subsection \ref{ssect:accelnow}. This expression is linear in $\kappa$, hence for $z$ values at which the coefficient of $\kappa$ vanishes, the density {\it will} be independent of curvature {\it for given $d_{L}(z)$}. The condition for this to be realized can be found as
\begin{equation}
D^2=\left(\frac{dD}{dz}\right)^2 (z+1)^2.
\label{eq:z*cond}
\end{equation}
which means that the slope of $D$, written as function of $u=z+1$, is equal to the slope of the chord connecting the point to the origin (note that both $u$ and $D$ are positive in the region of interest). Since the function  $D(u)$ starts at the point (1,0) for the first quadrant, it must have at least one inflection point for eq.(\ref{eq:z*cond}) to be satisfied, and so it is not at all obvious that a solution will exist. The curvature-independence of density makes these points, --if and when they exist-- special, and our analysis suggest that one such point exist for our universe; we call it $z_{*}$.

This point can be useful. For example, assuming that all contributions to the energy density of the universe are positive, it enables us to put an upper limit on the current value of matter density of our universe by equating (\ref{rho_kappa_z}) to the density of matter-dominated universe at $z=z_*$:
\begin{equation}
\frac{3c^4}{8\pi G}\left(\frac{dD}{dz}\right)_{(z=z_*)}^{-2}=\rho_{\rm 0m,max} (z_*+1)^3;
\end{equation}
using eq.(\ref{eq:z*cond}) and the definition of $D(z)$,
\begin{equation}
\rho_{\rm 0m,max} = \frac{3c^4}{8\pi G}\frac{z_*+1}{d_{L}^{2}(z_{*})},
\end{equation}
which can be converted into an upper bound for the density parameter for matter today, $\Omega_{0m,{\rm max}}$. The results for the 15 ``natural''  $d_L(z)$ functions referred to after figure \ref{fig:grid-a-ddot-supernova-grb} are given in Table \ref{tablezstar}. It is worth noting that the newly introduced variable $y5$  gives remarkably more consistent values of $z_*$, hence $\Omega_{0m,{\rm max}}$.

\begin{table} [b!]
\caption{\textsf{$z_*$ and $\Omega_{0m,{\rm max}}$ with their estimated errors calculated for 15 different $d_L(z)$ functions and assuming Einstein gravity}.}
\centering

\begin{tabular}{| c | c | c | }
\hline
$d_L(z)$ & $z_*$  & $\Omega_{0m,{\rm max}}$  \\
\hline
y0-F2 & $1.34^{+0.04}_{-0.03}$ & $0.502^{+0.035}_{-0.040}$   \\
\hline
y0-F4 & $1.35^{+0.05}_{-0.03}$ & $0.498^{+0.038}_{-0.047}$   \\
\hline
y0-F6 & $1.38^{+0.05}_{-0.04}$ & $0.475^{+0.040}_{-0.042}$   \\
\hline
y0-F7 & $1.47^{+0.08}_{-0.06}$ & $0.421^{+0.046}_{-0.054}$   \\
\hline
y0-F8 & $1.43^{+0.07}_{-0.05}$ & $0.444^{+0.043}_{-0.049}$   \\
\hline
y5-F2 & $1.44^{+0.08}_{-0.05}$ & $0.438^{+0.044}_{-0.054}$   \\
\hline
y5-F4 & $1.46^{+0.09}_{-0.06}$ & $0.426^{+0.046}_{-0.054}$   \\
\hline
y5-F6 & $1.46^{+0.08}_{-0.06}$ & $0.425^{+0.047}_{-0.051}$   \\
\hline
y5-F7 & $1.46^{+0.09}_{-0.06}$ & $0.428^{+0.045}_{-0.054}$   \\
\hline
y5-F8 & $1.43^{+0.07}_{-0.05}$ & $0.441^{+0.041}_{-0.047}$   \\
\hline
y6-F2 & $1.34^{+0.04}_{-0.03}$ & $0.502^{+0.036}_{-0.040}$   \\
\hline
y6-F4 & $1.36^{+0.04}_{-0.04}$ & $0.491^{+0.044}_{-0.041}$   \\
\hline
y6-F6 & $1.38^{+0.05}_{-0.04}$ & $0.475^{+0.040}_{-0.042}$   \\
\hline
y6-F7 & $1.47^{+0.08}_{-0.06}$ & $0.421^{+0.046}_{-0.050}$   \\
\hline
y6-F8 & $1.43^{+0.07}_{-0.05}$ & $0.444^{+0.043}_{-0.049}$   \\
\hline
\end{tabular}

\label{tablezstar}
\end{table} 

For the total current density of the universe we have
\begin{equation}
\Omega_{0} = 1 + \frac{k c^{2}}{H_{0}^{2}a_{0}^{2}} = 1 + k' \kappa_{0} \frac{c^{2}}{H_{0}^{2}} \approx 1 + 0.18 \; k',
\end{equation}
where the first equality easily follows from (\ref{firsteinstein}), and the others from our definitions of $\kappa_{0}$ and $k'$. Since the density of radiation is negligible in the recent universe, the gap between these $\Omega_{0}$ values and the numbers in Table \ref{tablezstar} point to the existence of something else in the universe, the lower limit on whose current density is
\begin{equation}
\Omega_{0,\rm de} \geq 0.28. \label{eq:DE_limit}
\end{equation}

In principle, it is also possible to put upper limits on radiation energy density of the universe in a similar manner. However these limits will be about four orders of magnitude higher than the energy density of the cosmic background radiation, so they cannot meaningfully limit the number of relativistic species filling the universe. Considering the possibility of very large number of (sterile) massless particle species will give a smaller upper limit on the current density of matter+radiation, hence a larger lower limit on the density of {\it dark energy}. Hence we quote (\ref{eq:DE_limit}) as our result.

\begin{figure}[b!]
\caption{\textsf{The average $w(z)$, in the framework of Einstein gravity, found from the fifteen functions we used in Fig.\ref{fig:supernova-grb-avg-acc},  extended to negative $z$, i.e. to the future. The color coding has the same meaning as in  figures \ref{fig:a(t)graph}-\ref{fig:grid-a-ddot-supernova}.}}
\centering
\includegraphics[width=0.99 \columnwidth]{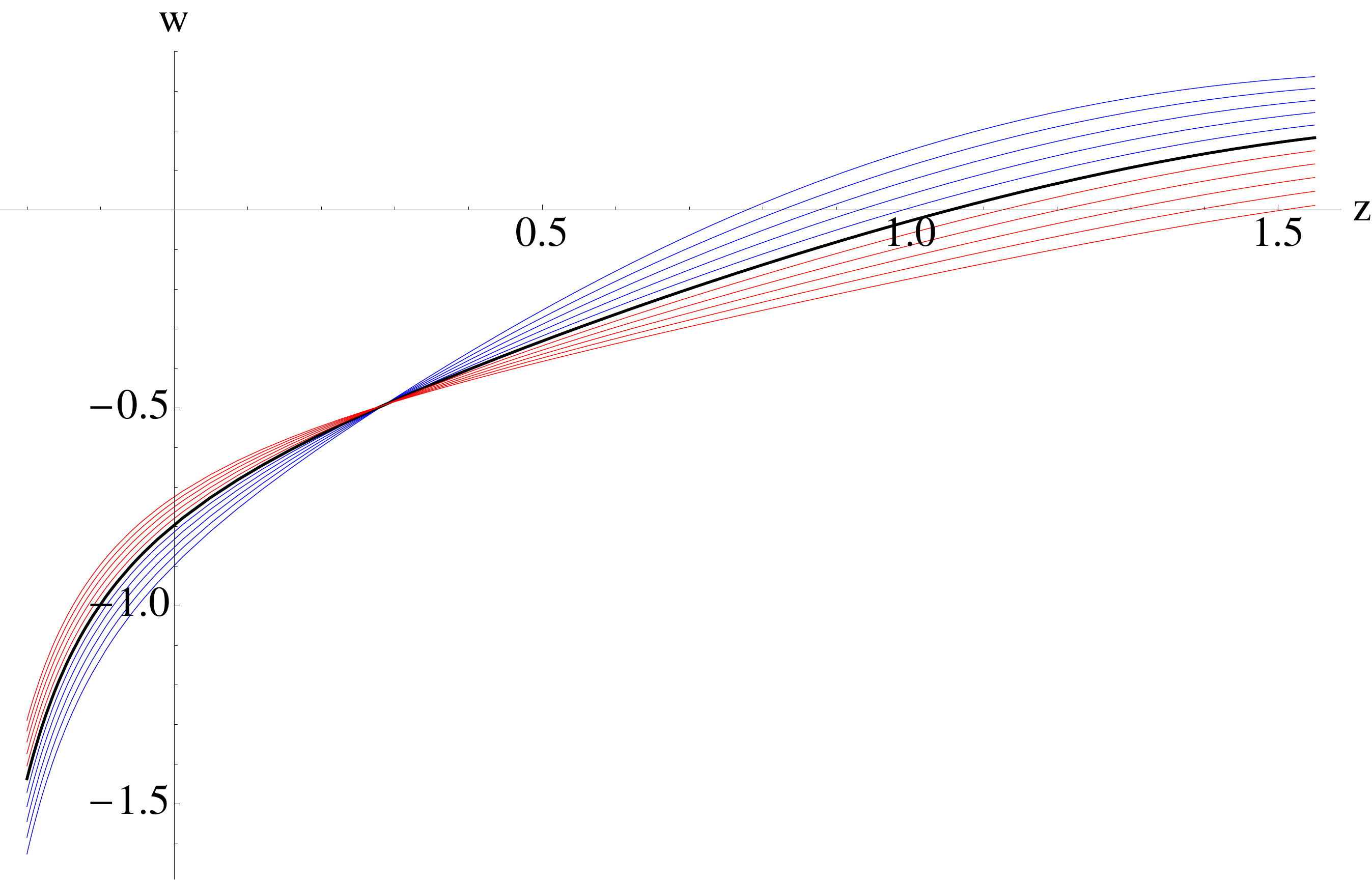}
\label{fig:eosparameter} 
\end{figure}

We can also use the second Einstein's equation which will give effective pressure $p(z)$ of the content of the Universe. Dividing it by the energy density we get $w(z)=p/\rho$; effective equation of state parameter for the Universe (Fig.\ref{fig:eosparameter})

Again we see an intersection point similar to the one in figure \ref{fig:rho_y5_f7}. With hindsight, we can analytically predict this point too, but it does not seem to have any physical significance. We find the current value of $w$ to be between -0.7 and -0.9 depending on the assumption for curvature, with an estimated error of 0.015. It should be noted that this is the total EoS parameter, not the $w$ of dark energy, which is reported in recent literature~\cite{cmb_first_peak, union21} with a central value less than -1; the two statements are not inconsistent. Extrapolation to the future however, shows a trend toward phantom-domination.

\section{Conclusion and Prospects}

We have fit various functions to the cosmological luminosity distance data, mostly the Union 2.1 Type Ia supernovae, and tried to extract information about the expansion history of the universe. The ans\"{a}tze used for the luminosity distance function $d_{L}(z)$ were motivated not by any physical model, but by facilitation of the operations for going from the function $d_{L}(z)$ to the function $a(t)$, the scale factor of the universe. We first took the number of parameters in the $d_{L}(z)$ ans\"{a}tze variable, forming families, and for each family we took the member with the minimum $\chi^{2}$/d.o.f. We also tried a number of different redshift variables.

Since in most of the work we did not use any law of gravity, our work is {\it model-independent}; in other words, we are simply studying the geometry of the universe, i.e. doing {\it cosmography}.

We found that in some models it does not make sense to try to extract more than two cosmological parameters, although this number is more  often three or four, up to six in some models. We found that Pade approximants and expressions involving $(z+1)$ as a factor give slightly better fits to $d_{L}(z)$ data than polynomials of the redshift variables, possibly multiplied by linear or quadratic exponentials of the same. Yet we found that no model fits significantly better than $\Lambda$CDM.

We demonstrated that {\it for ``measured'' $d_{L}(z)$ function}, the current acceleration of the universe does not depend on its curvature (a previously found, but not widely known result). But we found that the supernova data are too scattered and have too large errors to {\it by themselves} show a transition from deceleration to acceleration in the past of the universe, at least not for the ans\"{a}tze we used. We decided that the procedural problem lies in the convergence properties of polynomials, which form components of the ans\"{a}tze; and decided that it is possible to tame these divergences by including GRB data which reach out to much higher redshifts. Even though they are not really model-independent, we argue that due to their comparatively small number and large uncertainties, the work will not be contaminated much with physical models, while still serving to tame the dominance of higher order powers in polynomials. With their inclusion, we found the deceleration-acceleration transition at redshift $z_{\rm t, flat} = 0.50^{+0.08}_{-0.10}$ for a flat universe, larger (earlier) for positive spatial curvature, and smaller (later) for negative. 

Finally assuming Einstein gravity, we found that there is a special redshift value, call it $z_{*}$, at which the density of the universe is independent of its curvature, again {\it for ``measured'' $d_{L}(z)$ function}. This enables us to put an upper limit of roughly $0.50 \pm 0.04$ on the current density parameter of matter, dark or not, {\it using only the luminosity distance data}. This translates into a lower limit for current {\it dark energy} density parameter of roughly 0.3,  assuming the ``worst'' case for the curvature of the universe.

We can also find the total EoS parameter $w(z)$ for the universe; we find that the EoS is not yet {\it phantom}, but it seems to be evolving in that direction. Of course, assuming an alternative theory of gravitation would bring  (possibly qualitatively) different interpretations than in these last two paragraphs.

The coming decades should bring an increase in the number of the SNe Ia observations. If technology improves such that the uncertainties of the luminosity measurements decrease, it should be possible to find better fits; however, the scatter should also decrease for those fits to make more sense and the reconstruction of the scale factor function and other inferences made in this work to be more reliable. There is recent work \cite{SNdiv1,SNdiv2} suggesting that Type Ia supernovae might not be all identical after all, a potential development reminiscent of the discovery of the different types of Cepheids in 1940's. If confirmed, this would also lead to a reevaluation of existing data, possibly decreasing the scatter, and together with the new data, might lead to the required improvement.

\end{document}